\documentclass[10pt,
aps,
prd,
amsmath,
amssymb,
floats,
floatfix,
twocolumn,
superscriptaddress,
nofootinbib,
noshowpacs]{revtex4-2}

\usepackage[normalem]{ulem}
\usepackage{xfrac}
\usepackage{graphicx}
\usepackage{dcolumn}
\usepackage{bm}
\usepackage{xcolor}
\usepackage{hyperref}
\usepackage{fix-cm}

\definecolor{bluc}{cmyk}{1,0.9,0,0}
\definecolor{rossoCP3}{cmyk}{0,.88,.77,.40}
\definecolor{rosso}{cmyk}{0,1,1,0.4}
\definecolor{rossos}{cmyk}{0,1,1,0.55}
\definecolor{rossoc}{cmyk}{0,1,1,0.2}
\definecolor{verdes}{cmyk}{0.92,0,0.59,0.4}

\hypersetup{colorlinks, bookmarksopen, bookmarksnumbered, citecolor=verdes, linkcolor=bluc, pdfstartview=FitH, urlcolor=rossos}

\usepackage{multirow,booktabs}

\newcommand\redsout{\bgroup\markoverwith{\textcolor{red}{\rule[0.5ex]{2pt}{1pt}}}\ULon}

\begin{document}

\preprint{APS/123-QED}

\title{Signatures of cubic gravity in the strong regime}
\author{Flavio C. S\'anchez}
\affiliation{Unidad Acad\'emica de F\'isica, Universidad Aut\'onoma de Zacatecas, 98060, Zacatecas, M\'exico.}
\affiliation{Department of Physics, Faculty of Science and Engineering, Swansea University, Singleton Park,
SA2 8PP, Swansea, United Kingdom.}

\author{Armando A. Roque}
\author{Javier Chagoya}
\affiliation{Unidad Acad\'emica de F\'isica, Universidad Aut\'onoma de Zacatecas, 98060, Zacatecas, M\'exico.}%

\begin{abstract}
We investigate the effects of 
Einsteinian cubic gravity in the strong gravitational regime. In the first part, we explore analytical solutions for a static, spherically symmetric metric, establishing the existence of maximally symmetric de Sitter solutions, as well as asymptotically de Sitter solutions, with an effective cosmological constant. We also study, analytically and numerically, how the horizon properties are affected by cubic gravity. Our results reveal that a positive coupling constant reduces the horizon size, while a negative one increases it. In the second part, we analyze potential observational signatures of cubic terms, focusing on their effects on the bending of light. Specifically, we investigate the angular difference, related to the deflection angle but valid near the source, along with the behavior of the photon sphere. Our findings show that the strongest effects of the cubic terms occur in the strong gravity regime, and there exists a direct relationship between the value of the coupling constant and the photon sphere position, opening up the possibility to constrain cubic gravity with black hole shadows.
\end{abstract}

\maketitle

\section{Introduction}

Einsteinian cubic gravity is a higher-order gravity theory proposed by Pablo B. and Pablo A. C.~\cite{Bueno:2016xff}. This model incorporates a cubic term, constructed from contractions of the Riemann tensor, into the Einstein-Hilbert action with a constant $\Lambda$. On a maximally symmetric spacetime, there exists a unique combination $\mathcal{P}$ up to cubic order in curvature (in addition to Einstein gravity) that: (i) has dimension-independent couplings (at each order, the relative coefficients of the curvature invariants remain consistent across all space-time dimensions); (ii) shares the spectrum of Einstein's gravity (its linearized equations are equivalent, up to renormalization of Newton's constant) by propagating only a transverse and massless graviton on a maximally symmetric background; and (iii) is neither trivial nor topological in four dimensions. The inclusion of the cubic term in the action generally leads to field equations with derivatives higher than second order (we have up to six derivatives of the metric), which results in spacetime dynamics that differ from those of general relativity, as discussed in Ref.~\cite{Bueno:2016ypa}.

On an arbitrary four-dimensional spacetime, in addition to $\mathcal{P}$, there exist two other cubic densities, $\mathcal{C}$ and $\mathcal{C}'$ (see Eq.~$(2.16)$ in Ref.~\cite{Hennigar:2017ego} for the full expression), which preserve the properties mentioned above and provide a rich phenomenology. For instance, in a cosmological context where spacetime is described by the Friedmann-Lemaître-Robertson-Walker metric, the (unique) combination that maintains these properties is $\mathcal{P} - 8\mathcal{C}$. This theory is called Cosmological Einsteinian cubic gravity~\cite{Arciniega:2018fxj} and has been applied to describe the dynamics of inflation and late-time cosmological epochs~\cite{Arciniega:2018tnn, Erices:2019mkd, Cano:2020oaa, Quiros:2020uhr,Cisterna_2024}.\footnote{ In Ref.~\cite{Bueno:2016ypa}, the quartic extension of Einsteinian cubic gravity is presented, while Ref.~\cite{Cisterna:2018tgx} introduces and studies the quartic extension of Cosmological Einsteinian cubic gravity.}

For a static and spherically symmetric background in vacuum, the densities $\mathcal{C}$ and $\mathcal{C}'$ contribute only trivially to the field equations, leaving $\mathcal{P}$ as the unique non-trivial combination. In this context, cubic gravity admits the existence of black hole solutions where the temporal and radial metric components are inversely related~\cite{Bueno:2016lrh, Hennigar:2016gkm}, as in the Schwarzschild solution. However, the background geometry typically deviates from the Schwarzschild solution due to the inclusion of the cubic term, resulting in a rich additional phenomenology (e.g. Refs.~\cite{Hennigar:2018hza, Poshteh:2018wqy, Burger:2019wkq, Khodabakhshi:2020ddv, Gutierrez-Cano:2024oon}). As this work focuses solely on spherically symmetric vacuum solutions, the densities $\mathcal{C}$ and $\mathcal{C}'$ are not considered.

Like other higher-order gravitational theories, Einsteinian cubic gravity introduces new degrees of freedom compared to general relativity. In order to prevent their propagation and ensure that the theory retains the same spectrum as general relativity, these extra modes are considered infinitely heavy (by construction). However, this model exhibits unstable modes when perturbations over a static and spherically symmetric solution are considered~\cite{DeFelice:2023vmj, BeltranJimenez:2023mxp}. Similar pathologies arise in the cosmological version, with the presence of at least one ghost mode and a short-time-scale tachyonic instability in the regime of small anisotropies (see Ref.~\cite{Pookkillath:2020iqq} for details). In this sense, cubic gravity cannot be considered a complete theory but rather a perturbative correction (or effective low-energy theory) to general relativity. In other words, this model describes gravitational dynamics at low energies up to a certain energy scale, beyond which it exhibits Ostrogradsky instabilities. As discussed in Refs.~\cite{DeFelice:2023vmj, BeltranJimenez:2023mxp, Bueno:2023jtc}, for static and spherically symmetric spacetimes, the theory remains free of pathologies only within a specific regime where the Einsteinian cubic gravity is treated as an effective theory, $|\lambda\mathcal{P}/M_{\textrm{Pl}}^3|\ll |R|$.  Here, $R$ denotes the Ricci scalar, $M_{\textrm{Pl}}$ the reduced Planck mass, and $\lambda$ the coupling constant.\footnote{ The fact that our domain satisfies $|\lambda\mathcal{P}/M_{\textrm{Pl}}^3| \ll |R|$ implies that all non-perturbative effects arising from the cubic term, as well as quantum effects, are excluded. When these effects come into play, the breakdown of the effective theory approach leads to the previously mentioned problems~\cite{DeFelice:2023vmj, BeltranJimenez:2023mxp}.}

In this article, we focus on identifying potential signatures of Einsteinian cubic gravity by analyzing the bending of light in a triangular array of null geodesics, as proposed in Ref.~\cite{S_nchez_2023} (see Refs.~\cite{Takizawa:2020egm, Takizawa:2020dja, Arakida:2017hrm, Arakida:2020xil} for alternative proposals). This method defines an angular difference, $\alpha$, as the difference between the internal angles of triangular arrays constructed in different spacetimes. It can be used to extract contributions to $\alpha$ arising from modifications to general relativity. To achieve this, we compare triangular arrays constructed in Einsteinian cubic gravity with their counterparts in the general relativity framework, specifically using the Kottler~\cite{Kottler1918, weyl1919statischen} and Schwarzschild~\cite{Schwarzschild:1916uq, Schwarzschild:1916ae} solutions, since these are parametrically connected to the solutions in Einsteinian cubic gravity. In addition, we study the effect of cubic terms on the position of the photon sphere.

The paper is organized as follows: In Section~\ref{Sec.TS}, we first introduce the field equations that govern the dynamics of a static, spherically symmetric spacetime in Einsteinian cubic gravity. Then, we provide and discuss analytical and approximate solutions to this system. A family of maximally symmetric solutions is presented in Subsection~\ref{Sec.ExSol}, while the approximated iterative weak coupling, asymptotic, and near-horizon solutions are presented and discussed in Subsections~\ref{Sec.WCS},~\ref{Sec:AsymSol}, and~\ref{Sec:NHS}, respectively. These solutions extend those previously considered in Refs.~\cite{Bueno:2016lrh, Hennigar:2016gkm, Poshteh:2018wqy, Hennigar_2018}, motivated by similar mathematical considerations. In particular, we extend the analysis to both positive and negative coupling constant and to non-asymptotically flat spacetimes expressing all approximate solutions in terms of a single integration constant. To conclude this section, we discuss the horizon properties in Subsection~\ref{Sec.EvhPhS} and the validity regimes of the iterative solutions in Subsection~\ref{Sec.RegValit}. Possible signatures of the cubic terms, arising from their modification of light deflection, are addressed in Section~\ref{Sec2}. In Subsection~\ref{Sub.III.A}, the main ideas behind the methodology for calculating the angular difference are examined, while the results are presented in Subsection~\ref{Sec.III.B}. In Subsection~\ref{Sec.III.C}, we discuss the modifications that appear in the effective potential for a solution with the mass of SgrA$^{*}$ and their implications for the photon sphere position.

Finally, we conclude in Section~\ref{Sec.Concl} with an overview of our results. Some complementary material is presented in the appendixes.

\section{Static, spherically symmetric
solutions} \label{Sec.TS}

In four dimensions and using natural units, $\hbar=c=1$, Einsteinian cubic gravity is described by the action~\cite{Bueno:2016xff}
\begin{align}\label{Eq.AECG}
    S&=\int d^4x\sqrt{-g}\left[\frac{M_{\textrm{Pl}}^2}{2}\left(R -2\Lambda \right) + \frac{\lambda}{M_{\textrm{Pl}}^3}\mathcal{P}\right],
\end{align}
where $M_{\textrm{Pl}}:=1/\sqrt{8\pi G}$ is the reduced Planck mass, $R$ represents the Ricci scalar, and $\Lambda$ is a constant with units of energy squared, potentially related to the cosmological constant, often referred to as the bare cosmological constant. (The observed value of the cosmological constant in the theory at hand is obtained as a combination of $\Lambda$ and the contributions from the cubic terms. For a detailed discussion on the cosmological constant and its various contributions, see Ref.~\cite{Martin_2012}.) The parameter $\lambda$ is a coupling constant with units of energy and characterizes the energy scale at which the contributions of the cubic terms
\begin{align}\label{Eq.CubicTerm}
\mathcal{P}:=&12R_{\mu}{}^{\rho}{}_{\nu}{}^{\sigma}R_{\rho}{}^{\gamma}{}_{\sigma}{}^{\delta}R_{\gamma}{}^{\mu}{}_{\delta}{}^{\nu}+{R^{\rho \sigma}}_{\mu \nu}{R^{\gamma \delta}}_{\rho \sigma}{R^{\mu \nu}}_{\gamma \delta} \nonumber\\ 
&-12R_{\mu \nu \rho \sigma}R^{\mu \rho}R^{\nu \sigma}+8{R^\nu}_\mu {R^\rho}_\nu {R^\mu}_\rho,
\end{align}
are relevant. Notice that $\mathcal{P}$ is defined in terms of contractions of the Riemann tensor $R_{\mu\rho\nu\sigma}$. To avoid the pathologies discussed in the introduction, we focus exclusively on solutions valid in the regime where the radius $r_h$ of the black hole horizon satisfies $r_h \gg (\lambda/M_{\textrm{Pl}}^5)^{1/4}$, where cubic gravity can be treated as an effective theory.

The vacuum field equations (in their covariant form) are obtained by varying the action~(\ref{Eq.AECG}) with respect to the metric tensor $g^{\mu\nu}$, as follows:
\begin{align}\label{Eq.MotCov}
G_{\mu\nu}+g_{\mu\nu}\Lambda=&-\frac{\lambda}{2M_{\textrm{Pl}}^5}\mathcal{P}_{\mu\nu}.
\end{align}
Here, $G_{\mu\nu}$ is the Einstein tensor and $P_{\mu\nu}$, provided in Appendix~\ref{Ap.VCMEq}, represents the modification introduced by the cubic terms in Eq.~(\ref{Eq.CubicTerm}) to the vacuum covariant equations of general relativity.

For a static, spherically symmetric solution with a single metric function $f(r)$,
\begin{equation}\label{Eq.Metric}
    ds^2=-f(r)dt^2+\frac{dr^2}{f(r)}+r^2d\theta^2+r^2\sin^2\theta d\phi^2,
\end{equation}
the field equations~(\ref{Eq.MotCov}) reduce to one independent equation,
\begin{align}\label{Eq.Mot}
G&^{r}{}_{r}=\frac{M_{\textrm{Pl}}^2}{2}\left(\Lambda-\frac{1}{r^2}+\frac{f(r)}{r^2}+\frac{f'(r)}{r}\right)\nonumber\\
    &+\frac{6\lambda}{M_{\textrm{Pl}}^3}\bigg(g_{0} f'(r) + g_{1}f''(r) + g_{2} f'''(r)\bigg)\frac{f(r)}{r^2}=0,
\end{align}
where the functions $g_{i}$ are defined as:
\begin{subequations}\label{Eq.Components}
\begin{align}
    g_{0}:=&\bigg[\frac{4(1-f(r))}{r}+\left(4-\frac{1}{f(r)}\right)f'(r)\bigg]\frac{1}{r^2}, \\
    g_{1}:=&\frac{4\left(f(r)-1\right)}{r^2} - \frac{4 f'(r)}{r} + f''(r),  \\
    g_{2}:=&\frac{2(1-f(r))}{r} + f'(r).
\end{align}
\end{subequations}
The remaining components of the vacuum covariant equations~(\ref{Eq.MotCov}) for the line element~(\ref{Eq.Metric}) are either identically zero or equivalent to Eq.~\eqref{Eq.Mot}. It is crucial for our next results that Eq.~(\ref{Eq.Mot}) can be integrated to
\begin{align}\label{Eq.System2}
    \frac{2 C_0}{M_{\textrm{Pl}}^2 r}=& 1 - f(r)-\frac{\Lambda r^2}{3}-\frac{12\lambda}{M_{\textrm{Pl}}^5}\bigg[\frac{1}{3}\bigg(\frac{ 3f(r)f''(r)}{f'(r)^2}\nonumber\\
    &-1\bigg)f'(r)^2+\bigg(\frac{2f(r)f''(r)[1-f(r)]}{f'(r)^2}-1\bigg)\frac{f'(r)}{r}\nonumber\\
    &-\frac{2[1-f(r)]f(r)}{r^2}\bigg]\frac{f'(r)}{r}.
\end{align}
Here $C_0$ is an integration constant that for the case of an asymptotically flat black hole is the ADM mass~\cite{Bueno:2016lrh}.

In the next subsections, we identify a family of maximally symmetric solutions that exist within this theory and present a series of more general approximate solutions in three different scenarios. The first scenario represents a case where the main contribution to gravity arises from the general relativity component (weak coupling regime). The second involves determining the asymptotic solution. Finally, the third scenario corresponds to the strong gravity (high energy) regime near the horizon. In the final part of this section, we examine the existence and number of horizons and discuss the validity of the approximate solutions by comparing them with full numerical solutions.

\subsection{Maximally symmetric solutions}\label{Sec.ExSol}

We now show the existence of maximally symmetric solutions in the theory described by action~(\ref{Eq.AECG}). A maximally symmetric solution describes a spacetime that is homogeneous and isotropic. Such a spacetime satisfies the conditions that $R$ is a constant, $R_{\mu\nu}=g_{\mu\nu}R/4$, and $R_{\mu\nu\gamma\rho}=R(g_{\mu\gamma}g_{\nu\rho}-g_{\nu\gamma}g_{\mu\rho})$.

An Ansatz for the metric~(\ref{Eq.Metric}) that fulfills the aforementioned conditions is
\begin{align}\label{Eq.AnsMax}
f(r) = 1 - \Lambda_{\text{Eff}} r^2,
\end{align}
where $\Lambda_{\text{Eff}}$ is a constant with units of inverse energy squared. Substituting Eq.~(\ref{Eq.AnsMax}) into Eq.~(\ref{Eq.System2}), we arrive at
\begin{align}\label{Eq.vaccumsol}
2C_0 - \frac{16r^3}{M_{\textrm{Pl}}^3}\left(\Lambda_{\text{Eff}}^3\lambda + \frac{M_{\textrm{Pl}}^5}{16}\Lambda_{\text{Eff}} - \frac{M_{\textrm{Pl}}^5\Lambda}{48}\right) = 0.
\end{align}
Equation~(\ref{Eq.vaccumsol}) can be solved for all $r$ only if $C_0=0$, consistent with the fact that $C_0$ represents the ADM mass of the solution, which is not present in the Ansatz~(\ref{Eq.AnsMax}). In this case, Eq.~(\ref{Eq.vaccumsol})
reduces to a cubic algebraic equation for $\Lambda_{\text{Eff}}$ that admits real roots. Therefore, we obtain that a maximally symmetric solution is present in cubic gravity. Analogous to the de Sitter solution, the roots of $\Lambda_{\text{Eff}}$ act as an effective cosmological constant, which is given in terms of the bare cosmological constant $\Lambda$ and the coupling parameter $\lambda$. This modulation of the cosmological constant is known in the literature as self-tuning. Notice that the positive (negative) $\Lambda_{\text{Eff}}$ roots behave similarly to the de Sitter (anti-de Sitter) solution. A discussion of the real roots of Eq.~(\ref{Eq.vaccumsol}) is presented in Appendix~\ref{Ap.SecMSS}.

\subsection{Weak coupling solution}\label{Sec.WCS}

From Eq.~(\ref{Eq.System2}), in the limit where the cubic gravity contribution is non-dominant ($\lambda \ll 1$, weak coupling), it is natural to consider the metric function Ansatz
\begin{align}\label{Eq.AnsMax2}
f(r) = 1 - \frac{2 c_1}{r} - c_2 r^2,
\end{align}
where $c_1$ and $c_2$ are constants with units of inverse energy and inverse energy squared, respectively. However, after evaluating this Ansatz in Eq.~(\ref{Eq.System2}), the resulting equation becomes
\begin{align}\label{Eq.vaccumsol2}
    2(C_0- c_1 &M_{\textrm{Pl}}^2)-\frac{96 c_1 \lambda}{M_{\textrm{Pl}}^3}\bigg[c_2^2-\frac{7c_1}{r^3}\left(c_2-\frac{9}{14 r^2}+\frac{23c_1}{21 r^3}\right)\bigg]\nonumber\\[0.2cm]
    &-\frac{16r^3}{M_{\textrm{Pl}}^3}\left(c_2^3\lambda+\frac{M_{\textrm{Pl}}^5}{16}c_2-\frac{M_{\textrm{Pl}}^5\Lambda}{48}\right)=0,
\end{align}
and one can verify that vacuum solutions describing a black hole in empty spacetime with the structure in Eq.~(\ref{Eq.AnsMax2}) are not possible in this gravity model. Therefore, it becomes necessary to consider an approximate black hole solution with a more general structure for the metric function
\begin{equation}\label{eq:fr(r)weak}
    f_{\text{wcs}}(r) = 1-\frac{2C_0}{M_{\textrm{Pl}}^2 r}-\frac{\Lambda r^2}{3} + \sum_{n=1}^{m} \epsilon^n h_n(r).
\end{equation}
Here, the dimensionless order parameter $\epsilon:= \lambda \Lambda ^2/M_{\text{Pl}}^5$ is treated as small to ensure that we remain in the weak coupling regime. The contribution of cubic gravity in this scenario is encoded in the functions $h_n(r)$. Notice that this Ansatz is equivalent to decomposing the metric function $f(r)$ into the Kottler solution plus an expansion in terms of $h_n(r)$.

Inserting the Ansatz~(\ref{eq:fr(r)weak}) into Eq.~(\ref{Eq.System2}) and identifying the different orders in $\epsilon$, one arrives at a set of $m$ differential equations that can be solved iteratively for the functions $h_n$.\footnote{ A similar analysis can be carried out using the equation of motion~(\ref{Eq.Mot}). We refer the reader to Eq.~($2.15$) in Ref.~\cite{Hennigar:2016gkm} for the results. In this case, a set of integration constants, $C_{m}$, appears when solving the set of differential equations iteratively. These integration constants will be related to the constant $C_0$ through physical properties of the object, such as its horizon radius.} For the first three functions, we have
\begin{widetext}
\begin{subequations}\label{Eq.WcsCoeff}
\begin{align}
    h_1(r)=&\frac{32 C_0}{3 M_{\textrm{Pl}}^2 r}+\frac{16\Lambda r^2}{27}-\frac{224 C_{0}^{2}}{M_{\textrm{Pl}}^4 \Lambda r^4}\bigg(1-\frac{27}{14 r^2 \Lambda}+\frac{23 C_0}{7 M_{\textrm{Pl}}^2 r^3 \Lambda}\bigg),\\
    h_2(r)=&-\frac{2560 C_0}{27 M_{\textrm{Pl}}^2 r}-\frac{256 \Lambda r^2}{81}+\frac{35840 C_0^2}{9 M_{\textrm{Pl}}^4 \Lambda r^4}\bigg(1-\frac{243}{140 \Lambda r^2}-\frac{3441 C_0}{70 M_{\textrm{Pl}}^2\Lambda r^3}+\frac{21141 C_0}{70 M_{\textrm{Pl}}^2\Lambda^2 r^5}-\frac{43371 C_0^2}{70 M_{\textrm{Pl}}^4 \Lambda^2 r^6}-\frac{2187 C_0}{5  M_{\textrm{Pl}}^2 \Lambda^3 r^7}\nonumber\\
    &+\frac{124659 C_0^2}{70 M_{\textrm{Pl}}^4 \Lambda^3 r^8}-\frac{18009 C_0^3}{10M_{\textrm{Pl}}^6 \Lambda^3 r^9}\bigg),\\
    h_3(r)=&\frac{229376 C_0}{243 M_{\textrm{Pl}}^2 r}+\frac{16384 \Lambda r^2}{729}-\frac{1605632 C_0^2}{27  M_{\textrm{Pl}}^4 \Lambda r^4}\bigg(1-\frac{81}{49 \Lambda r^2}-\frac{20189 C_0}{196  M_{\textrm{Pl}}^2 \Lambda r^3}+\frac{28188 C_0}{49  M_{\textrm{Pl}}^2 \Lambda^2 r^5} + \frac{651261 C_0^2}{98 M_{\textrm{Pl}}^{4} \Lambda^2 r^6}+ H_3\bigg).
\end{align}
\end{subequations}
\end{widetext}
The term $H_3$ corresponds to the remaining terms, which are of order less than $\mathcal{O}(r^{-7}M_{\textrm{Pl}}^{-2}\Lambda^{-2})$, and its full expression is presented in Appendix~\ref{A.WCS}. For the interpretation of our results, the following observations are important: 
\begin{itemize}
    \item[i.]  The total mass of the object is the sum of all perturbative contributions, $M^{(wcs)}=C_0(1-\epsilon 16/3 + \epsilon^2 1280/27 -\epsilon^3 114688/243 + \dots)$. 
    \item[ii.] Similarly, the value of the cosmological constant is, $\Lambda^{(wcs)}=\Lambda(1 - \epsilon 16/9 + \epsilon^2 256/27-\epsilon^3 16384/243 + \dots)$.
\end{itemize}
Contrasting with the Kottler solution of general relativity, we see that cubic gravity redefines the mass and cosmological constant of the solution.

At this point, we have identified an iterative weak coupling solution $f_{\text{wcs}}$ Eq.~(\ref{eq:fr(r)weak}) with $m=3$ and $h_1, h_2, h_3$ given by Eqs.~(\ref{Eq.WcsCoeff}). Next, we proceed to identify an asymptotic solution that, as will be proven, is connected to the weak coupling solution in certain regimes.

\subsection{Asymptotic solution} \label{Sec:AsymSol}

\begin{table*}
\setlength{\tabcolsep}{10pt}
\begin{tabular}{l c c c c}
\toprule
    Range & $\lambda<-M_{\text{Pl}}^5/(12\Lambda^2) $ & $\lambda=-M_{\text{Pl}}^5/(12\Lambda^2) $	& $-M_{\text{Pl}}^5/(12\Lambda^2)<\lambda< 0$ & $\lambda>0 $\\ \midrule 
    $\Lambda_{\text{Eff}}$ & Eq.~(\ref{Eq.SolLamEff1}) & Eq.~(\ref{Eq.SolLamEff3}) & Eq.~(\ref{Eq.SolLamEff2}) & Eq.~(\ref{Eq.SolLamEff1}) \\ \midrule
    sign($\Lambda_{\text{Eff}}\Lambda$) &  $-$ & $+$ or $-$ & $+$ or $-$ & $+$ \\ \bottomrule
	\end{tabular}
    \caption{{\bf Summary of real-space solutions for the  \textbf{effective} constant $\Lambda_{\text{Eff}}$.} The regions where the sign of the cosmological constant $\Lambda$ matches the sign of effective constant $\Lambda_{\text{Eff}}$ are free from ghost instabilities. Notice that for $\lambda\geq-M_{\textrm{Pl}}^5/(12\Lambda^2)$, there exists at least one branch of solutions where this occurs.}\label{Table:Solpac}
\end{table*}

In order to identify the asymptotic solution of Eq.~(\ref{Eq.System2}), we assume the following Ansatz:
\begin{align}\label{Eq.AnsatAS}
    f_{\text{asy}}(r)=\sum_{n=0}^{m} a_{n} r^n + \frac{b_{n+1}}{r^{n+1}},
\end{align}
where $a_{n}$ and $b_{n+1}$ are constant coefficients. The terms proportional to $r^n$ include the expected de Sitter asymptotic behavior, while those proportional to $r^{-(n+1)}$ capture the asymptotic gravitational potential due to a massive source.

Evaluating Eq.~(\ref{Eq.AnsatAS}) in Eq.~(\ref{Eq.System2}) and considering the first ten terms (which require $m=11$ due to the differentiations involved), we find that, as expected, all coefficients $a_n$ are zero, with the exception of $a_0=1$ and $a_2=\Lambda_{\text{Eff}}$. The $b_n$ coefficients are determined iteratively, leading to the asymptotic solution
{\small
\begin{align}\label{Eq.SolAS}
    f_{\text{asy}}(r)=&1-\frac{\Lambda_{\text{Eff}}}{3} r^2-\frac{2C_0}{M_{\textrm{Pl}}^2 r}\left(\frac{1}{1+\frac{16\lambda \Lambda_{\text{Eff}}^2}{3 M_{\textrm{Pl}}^5}}\right)+\dots \,.
\end{align}}The ellipsis indicates the remaining terms, shown in  Eq.~(\ref{A.Eq.SolAS}) in Appendix~\ref{A.AS} and the effective cosmological constant $\Lambda_{\text{Eff}}$ is determined from the real root of the algebraic equation
\begin{align}\label{Eq.LambEff}
  \Lambda_{\text{Eff}}^3+\frac{9 M_{\textrm{Pl}}^5}{16\lambda}\Lambda_{\text{Eff}}-\frac{9 M_{\textrm{Pl}}^5 \Lambda}{16\lambda} =0.
\end{align}

At this point, we have two possible paths: the first (and the one used to construct numerical solutions in this work) involves fixing the value of $\Lambda_{\text{Eff}}$ to the observational cosmological constant, $1.1056\times 10^{-52} \text{m}^{-2}$~\cite{Planck:2018vyg}, and solving Eq.~(\ref{Eq.LambEff}) for $\Lambda$. This approach is based on the expectation that the observational value corresponds to $\Lambda_{\text{Eff}}$. The second path (followed in works such as Refs.~\cite{Bueno:2016lrh, Hennigar:2016gkm}) considers $\Lambda$ as the cosmological constant and computes $\Lambda_{\text{Eff}}$ from Eq.~(\ref{Eq.LambEff}). In this case, it is possible to use Cardano's formula, and from the sign of the discriminant $\triangle = \left(\frac{9 M_{\textrm{Pl}}^5 \Lambda}{32\lambda}\right)^2+\left(\frac{9M_{\textrm{Pl}}^5}{48\lambda}\right)^3$, identify the nature of the roots:

\noindent
\underline{\smash{$\triangle>0$}}: for $\lambda>0$ or $\lambda<-\frac{M_{\textrm{Pl}}^5}{12\Lambda^2}$ we have that $\triangle>0$ and Eq.~(\ref{Eq.LambEff}) has 
one real root given by:
\begin{align}\label{Eq.SolLamEff1}
\Lambda_{\text{Eff}}=\sqrt[3]{\frac{9 M_{\textrm{Pl}}^5 \Lambda}{32\lambda}+\sqrt{\triangle}}+\sqrt[3]{\frac{9 M_{\textrm{Pl}}^5 \Lambda}{32\lambda}-\sqrt{\triangle}}\,. 
\end{align}
It is straightforward to prove from Eq.~(\ref{Eq.SolLamEff1}) that the sign of the real root is opposite to that of the cosmological constant $\Lambda$ when $\lambda<-\frac{M_{\textrm{Pl}}^5}{12\Lambda^2}$ and matches it when $\lambda>0$ ensuring that this branch remains free of ghosts~\cite{Hennigar:2016gkm}. 

\noindent
\underline{\smash{$\triangle<0$}}: when $-\frac{M_{\textrm{Pl}}^5}{12\Lambda^2}<\lambda<0$, then $\triangle<0$ and all roots of Eq.~(\ref{Eq.LambEff}) are real and are given by,
\begin{align}\label{Eq.SolLamEff2}
\Lambda_{\text{Eff}}^{(k)}=\sqrt{\frac{3 M_{\textrm{Pl}}^5}{4|\lambda|}}\cos{\left(\frac{\theta+2k\pi}{3}\right)},
\end{align}
with $k=0, 1, 2$ and $0<\theta<\pi$, where $\theta=\cos^{-1}\left(\frac{9 M_{\textrm{Pl}}^5 \Lambda}{32\lambda}\bigg/\sqrt{\left(\frac{3M_{\textrm{Pl}}^5}{16|\lambda|}\right)^3}\right)$. In this case, at least one root shares the same sign as the cosmological constant.

\noindent
\underline{\smash{$\triangle=0$}}: finally, when $\lambda=-\frac{M_{\textrm{Pl}}^5}{12\Lambda^2}$, then $\triangle=0$, all roots are real, and two of them are identical:
\begin{align}\label{Eq.SolLamEff3}
\Lambda_{\text{Eff}}=\frac{3\Lambda}{2} \; \text{(double root)}, \quad \Lambda_{\text{Eff}}=-3\Lambda.
\end{align}
Like the previous case, we have one solution with the same sign as $\Lambda$.

In summary, when solving Eq.~(\ref{Eq.LambEff}), for $\lambda\geq -\frac{M_{\textrm{Pl}}^5}{12\Lambda^2}$ there will always be a \textit{non-problematic} solution (with the same sign as $\Lambda$) that ensures the asymptotic solution~(\ref{Eq.SolAS}) remains ghost-free. This result aligns with the reported in Ref.~\cite{Hennigar:2016gkm}, with the particular distinction that our approach is expressed solely in terms of the integration constant $C_0$. Table~\ref{Table:Solpac} illustrates the real-space solutions of Eq.~(\ref{Eq.LambEff}).

To conclude our analysis, it is important to note that if we consider $\lambda$ to be small, we can perform a power expansion of the asymptotic solution, Eq.~(\ref{Eq.SolAS}), in terms of the order parameter $\epsilon$ (defined in Section~\ref{Sec.WCS}) and recover the weak coupling solution, Eq.~(\ref{eq:fr(r)weak}). In this case, $\Lambda_{\text{eff}}\approx \Lambda(1-16/9\epsilon+256/27\epsilon^2+\dots)\,$. This result is validated by the findings shown in Fig.~\ref{fig:PositCoupl}, discussed in Section~\ref{Sec.RegValit}.

\subsection{Near Horizon solution} \label{Sec:NHS}

In this section, we proceed to identify the power series solution of Eq.~(\ref{Eq.System2}) near a horizon. Considering the Ansatz~\eqref{Eq.Metric} for the metric, the black hole horizon is defined as a surface of radius $r=r_{h}$ where $f(r_h) = 0$. Furthermore, we assume $f'(r_h)\geq 0$ at the (non-cosmological) horizon. This is justified, since we are interested in the {\it exterior} black hole horizon, where the metric function $f(r)$ is expected to change sign from positive (outside the horizon) to negative (inside the horizon). Notice, however, that we are allowing for $f(r)$ to have a minimum at $r_h$, since this is also compatible with $f(r_h)=0$ and $f(r>r_h)>0$. According to these constraints, we consider the Ansatz
\begin{equation} \label{eq:fNearHor}
    f_{\text{nhs}}(r)=\sum_{n=1}^{m} a_n (r-r_h)^n,
\end{equation}
where the constant coefficients $a_n$ are determined by evaluating Eq.~(\ref{eq:fNearHor}) in Eq.~(\ref{Eq.System2}) (expanded around $r_h$) and solving order by order in $\epsilon := r-r_h$, and it is required that $a_1\geq0$.

At order zero in Eq.~\eqref{Eq.System2}, the first coefficient of the series~\eqref{eq:fNearHor} can be determined as
\label{Eq.FirstOrders}
\begin{align}\label{Eq.aSol}
a_1 =\frac{2(1-r_h^2\Lambda)}{r_h\left(1 \pm \sqrt{1+\frac{48\lambda}{M_{\textrm{Pl}}^5 r_h^4}(r_h^2\Lambda -1)}\right)}.
\end{align}
Considering the solution branch that yields a finite value for $f_{\text{nhs}}'(r_h)$ in the limit $\lambda=0$, one can use the first-order results to obtain an equation for the horizon radius in terms of the constants $C_0, \lambda$, and $\Lambda$, as follows:
\begin{widetext}
\begin{align}\label{Eq.rhSol}
    \frac{2C_0}{M_{\textrm{Pl}}^2 r_h}=1-\frac{4}{3M_{\textrm{Pl}}^5 r_h^4}\left \{\frac{\left(1+\mathcal A\right)M_{\textrm{Pl}}^5\Lambda r_h^6 + 12\lambda(\Lambda r_h^2 -1)\left[5 + 2\Lambda^2 r_h^4 + 3\,\mathcal A-2\Lambda r_h^2 \left(2+\mathcal A\right)\right]}{\left(1+\mathcal A\right)^3} \right \},\\
    \nonumber
\end{align}
where $\mathcal A=\sqrt{1+\sfrac{48\lambda(\Lambda r_h^2 -1)}{(M_{\textrm{Pl}}^5 r_h^4)}}$. Going to the next order, $a_3$ is determined as:
\begin{align}
    a_3&=-\frac{2 a_2^2}{3a_1\left(1+\frac{2}{a_1 r_h}\right)}\left[1+\frac{2\left(\frac{3}{2r_h}+a_1\right)\left(\frac{a_1}{2 a_2 r_h}-1\right)}{a_2 r_h}\right]-\frac{M_{\textrm{Pl}}^5\Lambda r_h}{36\lambda a_1^2 (1+\frac{2}{a_1 r_h})}\left(1+\frac{a_1}{\Lambda r_h}+\frac{a_2}{\Lambda}\right).
\end{align}
\end{widetext}
Continuing this process, one can solve iteratively for each coefficient $a_{n}$. This process provides a solution $f_{\text{nhs}}$ with $a_2$ as the only free parameter. The value of $C_0$ (or $r_h$) is computed from Eq.~(\ref{Eq.rhSol}) for a fixed $r_h$ (or $C_0$). It is important to note that $a_2$ is proportional to the second derivative at $r_h$, i.e. $a_2=f''(r_h)/2$. 

Next, we proceed to discuss the existence, number, and properties of the horizons in the cubic gravity model. In particular, we have identified that the approximate solution~(\ref{eq:fNearHor}) is valid only within a constrained region.

\subsection{Horizons properties}\label{Sec.EvhPhS}

\begin{figure*}
	\centering
\includegraphics[width=0.6\linewidth]{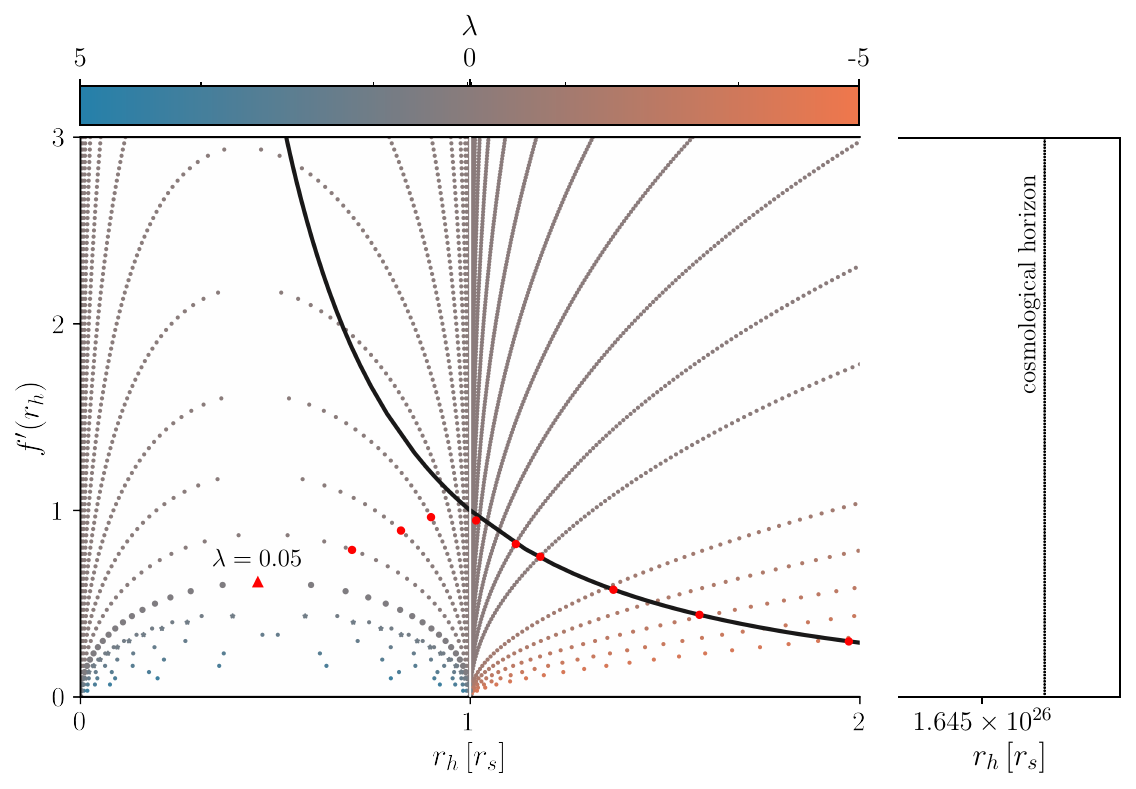}
\caption{{\bf Discrete solution space of Eq.~(\ref{Eq.Horiz}):} the main plot displays a set of solutions $(r_h, f'(r_h))$ for a range of $\lambda$ values, indicated by a color gradient. The black line denotes the solutions obtained in~\ref{Eq.FirstOrders}, while the red markers are some illustrative numerical solutions. In particular, the red triangle marker indicates the solution for $\lambda = 0.05$, reported in Fig.~\ref{fig:PositCoupl}. The secondary (right) plot illustrates the existence of a cosmological horizon across the entire solution space. A detailed discussion of this figure is provided in the main text. The results are expressed in the units defined in~(\ref{eq.ChangeVariables}).} \label{fig.HorProp}
\end{figure*}

In this work, we define a horizon as the radius $r_h$, corresponding to a real positive root of $f(r)$, where the first derivative is non-negative, i.e. $f(r_h) = 0$ and $f'(r_h) \geq 0$.

Focusing on the exterior (non-cosmological) black hole horizon, we observe that at $r_h$ the coefficients of $f''$ in Eq.~\eqref{Eq.System2} vanish, resulting in a cubic equation for the first derivative at the horizon, as follows:
\begin{align}\label{Eq.Horiz}
\frac{4\lambda}{M_{\textrm{Pl}}^3}f'(r_h)^{3} + & \frac{12\lambda}{M_{\textrm{Pl}}^3 r_h}f'(r_h)^2\\
&-\bigg(2C_0 - M_{\textrm{Pl}}^2r_h + \frac{M_{\textrm{Pl}}^2\Lambda r_h^3}{3}\bigg)=0.\nonumber
\end{align}
This equation determines the region within our parameter space where one or more event horizons may exist.

First, we consider the limiting case where $\Lambda=\lambda=0$. From Eq.~(\ref{Eq.Horiz}), it is easy to identify the existence of a single horizon radius, which corresponds to that of the Schwarzschild solution, i.e. $r_s=2C_0/M_{\textrm{Pl}}^2$.

When $\lambda=0$ and $\Lambda\neq 0$, the only non-zero term in Eq.~(\ref{Eq.Horiz}) is within the parentheses, corresponding to a cubic equation for $r_h$. In this case, the structure of the four possible cases described in Ref.~\cite{PhysRevD.15.2738} and presented in Ref.~\cite{Vandeev:2021yan} for Kottler spacetime is recovered. (In our notation, these regions are defined as follows: a single horizon when $\Lambda<0$ or $\Lambda=M_{\textrm{Pl}}^4/(9C_0^2)$; two horizons when $0<\Lambda<M_{\textrm{Pl}}^4/(9C_0^2)$; and no horizons when $\Lambda>M_{\textrm{Pl}}^4/(9C_0^2)$.)

Now, when only $\Lambda=0$, Eq.~(\ref{Eq.Horiz}) reduces to a quadratic equation for $r_h$, whose solutions can be written (after some manipulations) as
\begin{align}\label{Eq.rhL0Sol}
    &\frac{2C_0}{M_{\textrm{Pl}}^2 r_h} =1+\frac{12f'(r_h)^2}{C_0 M_{\textrm{Pl}}^3 r_h}\bigg(1+\frac{2C_0f'(r_h)}{3 M_{\textrm{Pl}}^2}\bigg)\lambda\nonumber\\
    &\quad \times\Bigg[1\pm\sqrt{1-\frac{12f'(r_h)^2 M_{\textrm{Pl}}^5\lambda}{(C_0M_{\textrm{Pl}}^3-2f'(r_h)^3\lambda)^2}}+\frac{2f'(r_h)^3\lambda}{C_0M_{\textrm{Pl}}^3}\nonumber\\
    &\quad \times \left(1\mp\sqrt{1-\frac{12f'(r_h)^2 M_{\textrm{Pl}}^5\lambda}{(C_0M_{\textrm{Pl}}^3-2f'(r_h)^3\lambda)^2}}\right)\Bigg]^{-1}.
\end{align}
From Eq.~(\ref{Eq.rhL0Sol}), one can infer that there exist solutions whose only horizon is equal to the Schwarzschild radius ($r_s = {2C_0}/{M_{\textrm{Pl}}^2}$), independently of the values of $\lambda$, these solutions have $f'(r_h) = 0$. Notice that the same conclusion can be obtained from Eq.~(\ref{Eq.Horiz}) with $f'(r_h) = 0$. The fact that Eq.~\eqref{Eq.System2} is well behaved when $f\to0$ and $f'\to0$ simultaneously suggest that the metric function is smooth at the horizon. In a recent work, Ref.~\cite{Wang:2024ehd}, the author presents a novel class of solutions featuring naked singularities with these characteristics. In particular, our results explain why, for all values of $\lambda$, the position of the critical horizon remains equal to the Schwarzschild radius. Continuing with the analysis for $\Lambda = 0$ and $f'(r_h)>0$, let us discuss the particularities of the cases $\lambda <0$ and $\lambda>0$:

\medskip
\noindent
\underline{$\lambda<0$}: the expression inside the brackets in Eq.~\eqref{Eq.rhL0Sol} is positive for the branch with the upper signs and negative for the other branch. (Notice that the square root is greater than 1.) It can be proven that there exists only one real positive solution $r_h$ which corresponds to the upper signs in Eq.~\eqref{Eq.rhL0Sol}.\footnote{The lower signs always lead to $r_h<0$, this can be shown by taking Eq.~\eqref{Eq.Horiz} with $\Lambda=0$ and rewriting it as a quadratic for $r_h$ where the coefficient of the quadratic term is positive and the coefficients of the linear and zero-order terms are negative. Thus, from the general quadratic formula, it is easy to verify that the solution with the lower sign is always negative.} In this case, the right-hand side of Eq.~(\ref{Eq.rhL0Sol}) is less than one, indicating a horizon radius greater than the Schwarzschild radius, i.e. $r_h(\lambda<0;\Lambda=0) > r_h(\lambda=0;\Lambda=0)$. This result is consistent with the findings of Ref.~\cite{Bueno:2016lrh}.

\medskip
\noindent
\underline{$\lambda>0$}: in this case, both roots $r_h$ of Eq.~(\ref{Eq.rhL0Sol}) are positive in the region determined by
\begin{align}
\lambda\leq \frac{M_{\textrm{Pl}}^5}{f'(r_h)^3}\left[\frac{C_0}{2M_{\textrm{Pl}}^2}+\frac{3}{2f'(r_h)}\left(1-\sqrt{1+\frac{2f'(r_h)C_0}{3M_{\textrm{Pl}}^2}}\right)\right].
\end{align}
In this region, the expression inside the brackets in  Eq.~(\ref{Eq.rhL0Sol}) is always positive, which implies that the right-hand side is greater than 1. This indicates that both solutions for $r_h$ are smaller than the Schwarzschild radius, i.e. $r_h(\lambda > 0; \Lambda = 0) < r_h(\lambda = 0; \Lambda = 0)$, for both the upper and lower signs in Eq.~(\ref{Eq.rhL0Sol}).

In the most general case, $\Lambda\neq0$ and $\lambda\neq0$, we need to study Eq.~(\ref{Eq.Horiz}) numerically. Solving Eq.~\eqref{Eq.LambEff} for $\Lambda$ we find  $\Lambda=\Lambda_{\text{Eff}}+{16\lambda \Lambda_{\text{Eff}}^3}/{9M_{\textrm{Pl}}^5}$. To ensure the desired asymptotic behavior, $\Lambda_{\text{Eff}}$ is fixed to the observational value of the cosmological constant. Our results indicate the existence of at least one event horizon corresponding to \textit{the cosmological horizon} both for $\lambda>0$ and $\lambda<0$. This cosmological horizon is indicated with a vertical black line in the secondary plot of Fig.~\ref{fig.HorProp}. Similar to the case $\Lambda=0$, for $\lambda<0$ (with $f'(r_h)>0$), all possible horizon radii (potentially more than two in some regions) are greater than the Schwarzschild radius. In contrast, for $\lambda > 0$, each of the horizons (except the cosmological horizon) is smaller than the Schwarzschild radius.

Figure~\ref{fig.HorProp} shows a discrete sample of the solution space of Eq.~(\ref{Eq.Horiz}), constrained to $r_h\geq0$ and $f'(r_h)\geq0$ within the range $-5\leq\lambda\leq5$. It is important to note that not all roots (represented by circular markers) exhibit the expected asymptotic behavior. For instance, when $\lambda=0.05$, the only root that satisfies this condition is denoted with a triangular marker. (The complete numerical solution is shown in Fig.~\ref{fig:PositCoupl}.) A potential method to identify roots with the desired asymptotic behavior is to use the near-horizon approximated solution, specifically the system~(\ref{Eq.FirstOrders}). The roots of this system are represented by a black line. Note that this approach is valid only for negative coupling. In this case, the black line aligns with the red markers, which represent the full numerical solutions. However, for the positive branch, this alignment does not occur, and the near-horizon approximated solution cannot be used as a seed.\footnote{ This is because, at $r_h$, the first and second derivatives are of the same order, which is not consistent with the assumed condition for the near-horizon approximated solution.} Instead, the asymptotic approximated solution must be used as the seed for full numerical integration. Additionally, as observed, when $|\lambda|$ increases, the value of $f'(r_h)$ for the solution with the expected asymptotic behavior (red markers) decreases. As expected, when $|\lambda|\to 0$, $r_h\to r_s$.

\subsection{Regimes of validity}\label{Sec.RegValit}

\begin{figure*}
    \centering
    \includegraphics[width=0.9\linewidth]{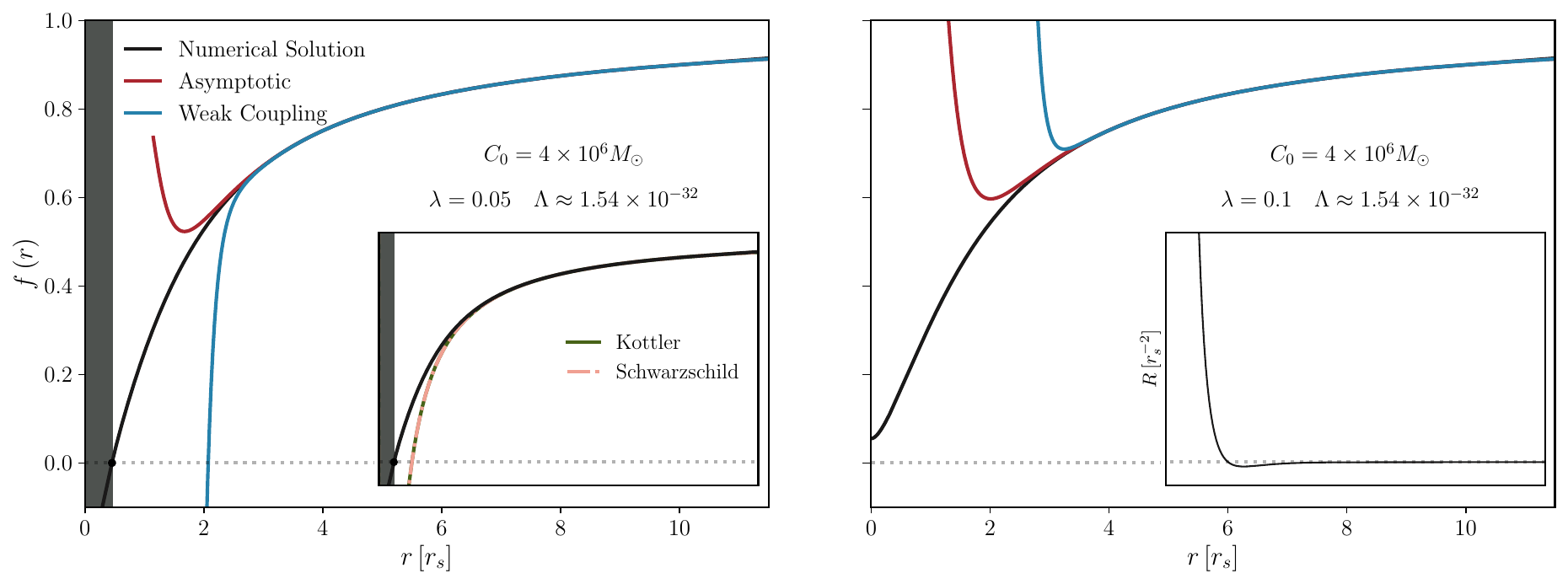}
    \caption{\textbf{Solutions with a positive coupling constant.} Numerical solutions (black lines) are shown for constants $C_0$ fixed to the mass of SgrA$^{*}$ and $\Lambda_{\text{Eff}} = \Lambda_{\text{CDM}}$, where the value of $\Lambda$ is determined using Eq.~(\ref{Eq.LambEff}). The solutions correspond to two values of the coupling constant: $\lambda = 0.05 [M_{\text{Pl}}^5 r_s^4]$ (left panel) and $\lambda = 0.1 [M_{\text{Pl}}^5 r_s^4]$ (right panel). The blue and red solid lines represent the asymptotic solution Eq.~(\ref{Eq.SolAS}) and the weak coupling solution Eq.~(\ref{eq:fr(r)weak}), respectively. As expected, the former agrees well for large values of $r$, while the latter is applicable for small values of the coupling constant $\lambda$. The inset in the left panel compares our numerical solution with the Kottler and Schwarzschild metrics, revealing a smaller event horizon than those predicted by these solutions. The right panel highlights the existence of solutions with a naked singularity, as demonstrated by the inset, which illustrates the divergence of the Ricci scalar at $r=0$.}\label{fig:PositCoupl}
\end{figure*}

\begin{figure*} 
	\centering
\includegraphics[width=0.9\linewidth]{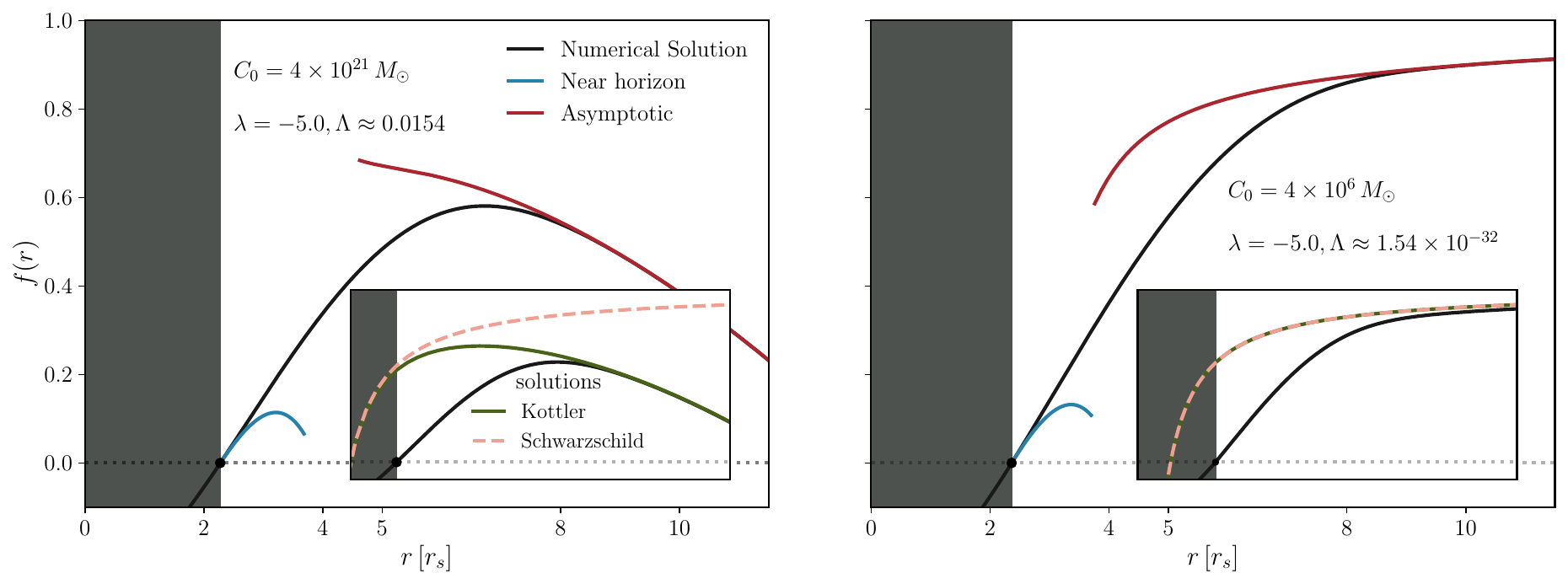}
\caption{\textbf{Solutions with a negative coupling constant.} Same as in Fig.~\ref{fig:PositCoupl}, but for the case of a negative coupling constant $\lambda=-5.0 \, [M_{\text{Pl}}^5 r_s^4]$. Compared to the positive case, the main difference is that the numerical integration is now started at the horizon. (Details are provided in the main text.) Notice that the solutions now exhibit a larger value for the event horizon than the Kottler and Schwarzschild solutions, as shown in the respective insets. The near-horizon solution Eq.~(\ref{eq:fNearHor}) (blue lines) is valid only within a small region near the horizon.}\label{fig:f(r)NumericalSolutions}
\end{figure*}

In previous sections, we presented and discussed the approximated solutions~(\ref{eq:fr(r)weak}),~(\ref{Eq.SolAS}), and~(\ref{eq:fNearHor}). Here, we analyze the validity of these solutions by comparing them with the full numerical solutions. To facilitate this comparison, we introduce the following dimensionless variables,
\begin{align} \label{eq.ChangeVariables}
    r:=\frac{2C_0}{M_{\textrm{Pl}}^2}\bar{r}, \quad \lambda:=\frac{16 C_0^{4}}{M_{\textrm{Pl}}^3}\bar{\lambda}, \quad \Lambda:=\frac{M_{\textrm{Pl}}^4}{4C_0^2}\bar{\Lambda},
\end{align}
into Eq.~(\ref{Eq.System2}). Notice that for $C_0=1/2$ and $M_{\text{Pl}}=1$ the new and old variables coincide. For simplicity, we omit the bars in the notation throughout the rest of the paper and when dimensionful quantities are required, they are specified explicitly.

To solve numerically the differential equation~(\ref{Eq.System2}), we need two appropriate boundary conditions. These can be established using the approximate series solutions Eq.~(\ref{eq:fr(r)weak}), Eq.~(\ref{Eq.SolAS}) or Eq.~(\ref{eq:fNearHor}), depending on the sign of $\lambda$. The strategy for the positive coupling $\lambda>0$ is to develop the numerical solution by starting at a point $r$ sufficiently far from the gravitational source and then, integrating toward the event horizon $r_h$. In this case, Eq.~(\ref{eq:fr(r)weak}) or Eq.~(\ref{Eq.SolAS}) (for weak coupling) is used to determine the value of $f$ and its first derivatives at this radius. For negative coupling $\lambda<0$, Eq.~(\ref{Eq.System2}) exhibits stiffness, requiring a different approach (the right-to-left integration becomes highly challenging). To address this, we use the iterative solution Eq.~(\ref{eq:fNearHor}) as a seed, carefully selecting the value of $a_2$ to ensure that the asymptotic solution is described by Eq.~(\ref{Eq.SolAS}).

Figure~\ref{fig:PositCoupl} presents the numerical solutions for two possible values of the positive coupling constant: $\lambda=0.05$ (left panel) and $\lambda=0.1$ (right panel). We used an adaptive explicit fifth-order Runge-Kutta routine for the numerical integration, with errors estimated using a fourth-order routine \cite{Virtanen_2020, DORMAND198019, Shampine1986SomePR}. To ensure that the solutions are ghost-free (see Section~\ref{Sec:AsymSol} for a discussion) and reproduce the expected asymptotic behavior, the value of $\Lambda_{\text{Eff}}$ is fixed to the observational cosmological constant $\Lambda_{\text{CDM}}$. The constant $C_0$ is taken as the mass of SgrA$^{*}$, $C_0=4\times 10^{6} \, M_\odot$~\cite{2016ApJ...830...17B}. As shown, the approximate series solutions~(\ref{eq:fr(r)weak}) (up to $\mathcal{O}(\epsilon^4)$) and~(\ref{Eq.SolAS}) (up to $\mathcal{O}(\lambda^2,C_0^3/r^9)$) agree with the numerical results for large values of $r$. In both cases, the validity of the series deteriorates as they approach the horizon, a behavior consistent with the assumptions used in their construction. However, it can also be observed that, as expected, the asymptotic series exhibits a wider range of validity compared to the weak coupling as the $\lambda$ value increases. On the other hand, for comparison, the corresponding Schwarzschild solution (with $M=C_0$) and Kottler solution (with $M=C_0, \Lambda=\Lambda_{\text{CDM}}$) are shown in the inset of the left panel. As noted and predicted in the previous section (for $\lambda > 0$), the non-cosmological horizon is smaller than those of the Schwarzschild and Kottler solutions. In the right panel of Fig.~\ref{fig:PositCoupl}, an illustrative solution with only one horizon (the cosmological horizon) is shown. Although the profile is regular at the origin, this solution contains a naked singularity, as the Ricci scalar diverges at $r=0$ (see the inset). Our numerical results indicate the existence of this type of solution when $\lambda\gtrapprox 0.09$ (with $C_0, \Lambda_{\text{Eff}}$ fixed to the values specified in the figure). Finally, the approximate near horizon solution Eq.~(\ref{eq:fNearHor}) is not shown because it is not valid in this branch (see Fig.~\ref{fig.HorProp}).

Figure~\ref{fig:f(r)NumericalSolutions} presents a similar analysis (again fixing $\Lambda_{\text{eff}}$ to the cosmological constant value), but this time for a negative coupling constant $\lambda = -5$. The left panel shows an illustrative (non-physical) solution to demonstrate the methodology used: connecting the near and asymptotic regions by the numerical solution (black line). The region near the horizon is described by the solution Eq.~(\ref{eq:fNearHor}) (blue line), while for large $r$, the solution matches the asymptotic approximation Eq.~(\ref{Eq.SolAS}) (red line). Our results indicate (and validate our analytical findings) that, for this branch, the exterior horizons are larger than those of the respective Schwarzschild and Kottler solutions (see the corresponding insets). The right panel shows the results obtained by fixing the constant $C_0$ to the mass of SgrA$^{*}$. 

It is important to note that the mass correction associated with the cubic terms is very small within the parameter space explored in this work. (For instance, based on the asymptotic solution in Eq.~(\ref{Eq.SolAS}) and using the parameters from the previous examples, the correction is of the order of $10^{-60}C_0$.) However, as discussed in the next section, the cubic terms lead to effects in the spacetime that can still be identified in certain regions (close to the source), despite the small magnitude of the mass correction.

In summary, for large values of $r$, the asymptotic and weak iterative solutions accurately describe the full numerical solutions in both branches. However, the near-horizon iterative solution is valid only for a negative coupling constant and within a small region around the horizon. In the next section, we combine the numerical and asymptotic solutions to construct triangular configurations and identify possible signatures by computing the angular difference, as presented in Ref.~\cite{S_nchez_2023}.

\section{Possible signatures of the cubic contribution in light deflection}\label{Sec2}
\begin{figure*}
    \centering
    \includegraphics[width=1\linewidth]{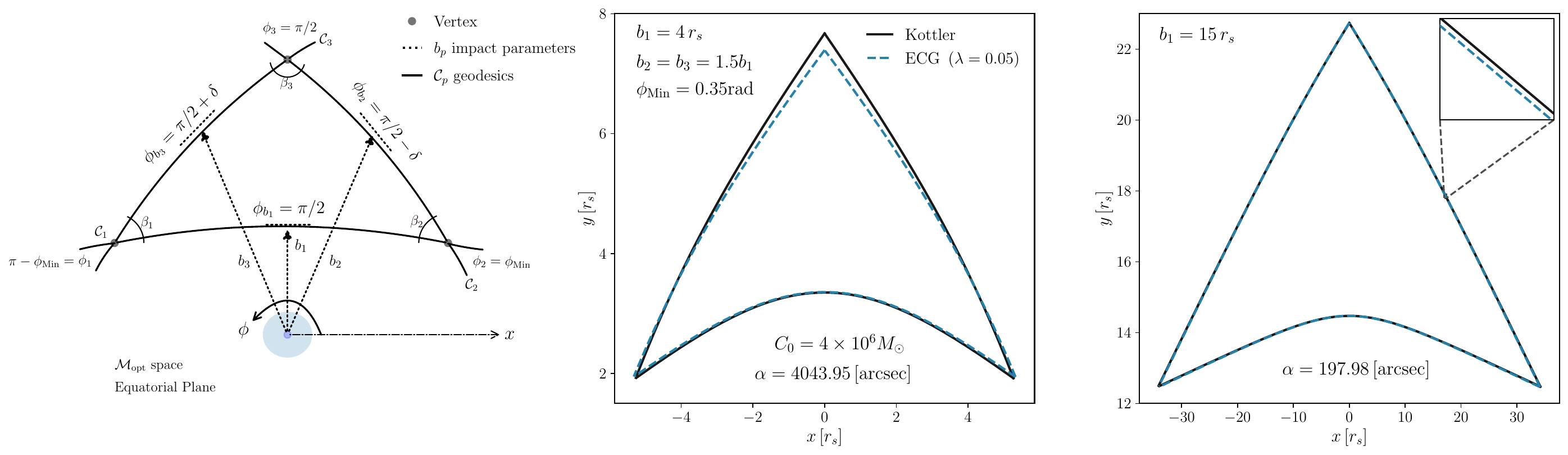}
    \caption{\textbf{Schematic and illustrative null geodesic triangular array.} The left panel displays a schematic representation of a triangular array with vertices $\phi_p$, bounded by three null geodesics $\mathcal{C}_p$ ($p = 1, 2, 3$) localized in the equatorial plane of a Riemannian space $\mathcal{M}_{\text{opt}}$ defined by the optical metric. Each geodesic $\mathcal{C}_p$ is associated with an impact parameter $b_p$ at an angular position $\phi_{b_p}$. For all triangular arrays explored in this paper, the vertices are fixed at: $\phi_1=\pi-\phi_{\text{Min}}, \phi_2=\phi_{\text{Min}}$, and $\phi_3=\pi/2$, where $\phi_{\text{Min}}=0.35$rad. The corresponding internal angles $\beta_p$, are used to compute the angular difference Eq.~(\ref{Eq.DeflecAng}). The central and right panels display triangle arrays where the spacetime under study is given by the numerical solutions of Eq.~(\ref{Eq.System2}) with a positive coupling constant $\lambda = 0.05$ (blue dashed line), and the background is described by the Kottler solution (solid black line). The central panel illustrates how the triangular array becomes more deformed due to its proximity to the gravitational source. In contrast, the right panel shows the same triangular array positioned much farther from the gravitational source. The insets reveal that, although these triangles appear similar, they exhibit minor discrepancies attributable to the $\lambda$ effect. Further details and discussions are provided in the main text.
 }\label{fig:Triangles}
\end{figure*}

In this section, we examine the angular difference $\alpha$, as proposed in Ref.~\cite{S_nchez_2023}, within the framework of Einsteinian cubic gravity. As mentioned in the introduction, this angular difference provides an alternative method for determining the deflection angle in static, spherically symmetric, and non-asymptotically flat spacetimes~\cite{Arakida_2021}. The method involves calculating the internal angles of a closed polygon (specifically, a triangle) formed by null geodesics in the spacetime of interest, which encode the effects of local curvature, and comparing them to a reference polygon (triangle) constructed in a background spacetime. A schematic representation of the construction of this triangle is shown in the left panel of Fig.~\ref{fig:Triangles}.

As derived in Ref.~\cite{S_nchez_2023}, the angle $\alpha$ is determined from the angular difference between the internal angles of the triangular array, as follows:
\begin{align}\label{Eq.DeflecAng}
    \alpha = \left| \sum_{p=1}^3 \left( \beta_p^{(1)} - \beta_p^{(2)} \right) \right|, 
\end{align}
where $\beta_p^{(i)}$ are the internal angles of the triangle placed in the spacetimes: $i=1$ the spacetime of interest 
and $i=2$ the spacetime that serves as background. For instance, if $i=1$ is the Schwarzschild spacetime and $i=2$ is Minkowski, then in the limit where the impact parameters of the null geodesics $\mathcal{C}_2$ and $\mathcal{C}_3$ (see Fig.~\ref{fig:Triangles}) become sufficiently large (i.e., $b_2 = b_3 \to \infty$), the textbook formula for the deflection angle of the Schwarzschild metric is recovered.\footnote{ To review the procedure for obtaining the textbook formula for the deflection angle of the Schwarzschild metric, we refer the reader to Ref.~\cite{Misner:1973prb}, Section $7.1$ of Ref.~\cite{will_2018}, or Section $5$ of Ref.~\cite{Weinberg:1972kfs}.} That is, the triangular configuration behaves as if the light source and the observer were located in the flat region at spatial infinity~\cite{S_nchez_2023}.

Setting aside the fact that the observed spacetime is not asymptotically flat, the traditional method for computing the deflection angle captures only combined information from all sources that curve spacetime. This makes it impossible (or highly complex) to disentangle and characterize individual contributions, such as those arising from modifications to general relativity or spacetime's non-flatness. In contrast, using the angular difference, it is possible to determine the contribution of different sources by comparing them with an appropriate background. For instance, the contribution of an object with mass $M$ in non-asymptotically flat spacetime (e.g. $i=1$, Kottler) can be determined when the background being studied contains no mass source (e.g. $i=2$, de Sitter). The measurement of angular differences could be particularly promising for exploring modified gravity scenarios and may be applied in future space missions, such as LATOR~\cite{Turyshev:2003wt}, ASTROD-GW~\cite{Ni:2012eh}, or LISA~\cite{Bayle:2022hvs, Amaro_Seoane_2023}, by deploying a triangular array of satellites connected via laser baselines.

\subsection{Triangular numerical construction}\label{Sub.III.A}

To set up the triangular configurations, we extend the methodology presented in Ref.~\cite{S_nchez_2023}, incorporating a numerical metric profile as the metric component. Below, we outline the basic ideas of our methodology and refer the reader to Ref.~\cite{S_nchez_2023} for additional details.\footnote{ Our numerical codes are publicly available in~\cite{flavio_repository_2025} both in Wolfram Mathematica and Python.}

The first step in constructing a triangular array is to determine the equation for the null geodesics $\mathcal{C}_p$ that outline the array. From the null condition $ds^2 = 0$, we have that for metrics~(\ref{Eq.Metric}), the equation for the trajectory of null geodesics reduces to
\begin{align}\label{Eq:Orbtray}
    \left(\frac{dr}{d\phi}\right)^2+f(r) r^2=\frac{r^4}{b^2},
\end{align}
where $b:=L/E=[r^2/f(r)]d\phi/dt$ is the impact parameter, defined in terms of the conserved photon energy $E=f(r)dt/d\lambda$ and angular momentum $L=r^2d\phi/d\lambda$, with $\lambda$ an affine parameter along the light trajectory. In terms of the dimensionless variables~(\ref{eq.ChangeVariables}) and introducing the substitution $u:=1/r$, Eq.~(\ref{Eq:Orbtray}) can be rewritten as
\begin{align}\label{Eq:OrbtrayU}
    \left(\frac{du}{d\phi}\right)^2+f(u) u^2=\frac{1}{b^2},
\end{align}
where, as before, we omit the bar over the dimensionless variables. The asymptotic (or weak coupling) solutions are used to extend the numerical results to larger values of $r$.

The second step consists in solving Eq.~(\ref{Eq:OrbtrayU}) with appropriate boundary conditions for the three null geodesics $C_1, C_2$, and $C_3$, such that a triangular configuration is formed. For simplicity, the vertices are chosen as
\begin{align}
    \phi_{1}=\pi-\phi_{\text{Min}},\quad \phi_2=\phi_{\text{Min}},\quad \phi_{3}=\pi/2, \nonumber
\end{align}
as shown in the left panel of Fig.~\ref{fig:Triangles}. For the geodesic $\mathcal{C}_1$, we apply a Neumann boundary condition, $du/d\phi=0$ at $\phi=\pi/2$. For the remaining geodesics, $\mathcal{C}_2$ and $\mathcal{C}_3$, we use Dirichlet boundary conditions such that the bottom vertex of $\mathcal{C}_2$ ($\mathcal{C}_3$) coincides with the right (left) vertex of $\mathcal{C}_1$ at $\phi_2=\phi_{\text{Min}}$ ($\phi_3=\phi_{\text{Max}}$), where $[\phi_{\text{Min}}, \phi_{\text{Max}}]$ is the angular domain of integration for $\mathcal C_1$~\footnote{ The choice of these boundary conditions is meant to construct a simple approximation of the geodesics. Thus, the geodesic $C_1$ determines the boundary conditions for the other two, and its own boundary condition depends on the physical scenario being described (e.g. impact parameter, compact object).}. For simplicity, the impact parameters $b_2$ and $b_3$ of geodesics $\mathcal{C}_2$ and $\mathcal{C}_3$ respectively, are taken to be equal and fixed to $1.5 b_1$. We only vary the parameter $b_1$, which corresponds to the geodesic $\mathcal{C}_1$ that describes the closest trajectory to the gravitational source. The numerical integration is carried out using the same methodology as in Section~\ref{Sec.RegValit}. 

The central and right panels of Fig.~\ref{fig:Triangles} show two examples of triangular arrays constructed using the methodology described above. In this case, the spacetime of interest ($i = 1$) corresponds to our numerical solutions to Eq.~(\ref{Eq.System2}) with a positive coupling constant $\lambda = 0.05$ (blue dashed line), and the background considered ($i = 2$) is the Kottler solution (solid black line). As expected, when $b_1$ is smaller (closer to the gravitational source), the triangle becomes significantly deformed (see the center panel). In contrast, when $b_1$ is large, the triangle retains a more regular shape (see the right panel). In both cases, $C_0$ (the source for the Kottler solution and the integration constant for the cubic solution) is fixed to the mass of SgrA$^{*}$.

Once the triangles are constructed, we compute their internal angles $\beta_p$ and, from these, the angular difference $\alpha$ as the final step. The internal angles are obtained from Ref.~\cite{S_nchez_2023}, as follows:
\begin{subequations}\label{EqAngRel}
\begin{align}
\beta_1&:=\psi_{\phi_{\text{Max}}}^{(3)}-\psi_{\phi_{\text{Max}}}^{(1)},\\
\beta_2&:=\psi_{\phi_{\text{Min}}}^{(1)}-\psi_{\phi_{\text{Min}}}^{(2)},\\
\beta_3&:=\psi_{\pi/2}^{(2)}-\psi_{\pi/2}^{(3)},
\end{align}
\end{subequations}
where $\psi_a^{(p)}$ indicates an angle computed for the geodesic $p$ at $\phi=a$, using the tangent formula~\cite{Perlick:2021aok}:
\begin{align}
\tan \left(\psi_{a}^{(p)}\right)&=\sqrt{\frac{\bar{g}_{\phi\phi}(r_p)}{\bar{g}_{rr}(r_p)}}\frac{d\phi}{dr_p}\bigg|_{\phi=a}. \label{EqTang}
\end{align}
The radial derivative is calculated using second-order accurate central differences~\cite{Alfio2007}. Finally, applying the angular formula~(\ref{Eq.DeflecAng}), we compute the angular difference $\alpha$ for the spacetime solution under study.

\subsection{Cubic terms contribution}\label{Sec.III.B}

\begin{table}[t]
 \renewcommand{\tabcolsep}{9pt}
 \centering
 \caption{\textbf{Contribution of the cubic term to the angular difference $\alpha$.} Triangular arrays with impact parameters $b_1\in [10,10^4]$ 
 (rows), where the spacetimes of interest are solutions to Einsteinian cubic gravity with different values of the coupling constant $\lambda$ 
 (columns), satisfying $|\lambda\mathcal{P}/M_{\textrm{Pl}}^3| \ll |R|$ within the explored regions. 
 Up to our numerical resolution, the angular difference using either Schwarzschild or Kottler as background remains the same. The complementary parameters associated with these triangular arrays are shown at the bottom of the table. \label{tabla_1}}
 \begin{tabular}{l@{\hskip .02 in}  c@{\hskip .08 in} c@{\hskip .08 in} c@{\hskip .08 in} c@{\hskip .08 in}}
     \toprule
     &\multicolumn{4}{c}{$\alpha$ [arcsec]}\\
     \cmidrule{2-5}
     &\multicolumn{4}{c}{$\lambda\; \left[\sfrac{16 C_0^{4}}{M_{\textrm{Pl}}^3}\right]$}\\[0.1cm]
     $b_1 [r_s]$ & $-5.0$ & $-1.5$ & $0.05$ & $0.1$\\
     \cmidrule{2-5}
     $10$ & $6.46 \times 10^{3}$ & $5.65\times10^3$ & $0.61 \times 10^{2}$ & $1.22 \times 10^{2}$\\
     $10^2$ & $4.28 \times 10^{-2}$ & $1.28 \times 10^{-2}$ & $4.28 \times 10^{-4}$ & $8.56 \times 10^{-4}$\\
     $10^3$ & $6.09 \times 10^{-5}$ & $4.13 \times 10^{-5}$ & $3.22 \times 10^{-6}$ & $3.22 \times 10^{-6}$\\
     $10^4$ & $2.04 \times 10^{-8}$ & $2.04 \times 10^{-8}$ & $2.04 \times 10^{-8}$ & $2.04 \times 10^{-8}$\\
 \bottomrule
 \end{tabular}
 \vspace{0.5em}
\noindent Config: $C_0 = 4.3 \times 10^6 M_\odot, b_2 = b_3 = 1.5b_1, \phi_{\min} = 0.3 \, \text{rad}$
 \end{table}

\begin{figure*}
    \centering
    \includegraphics[width=0.9\linewidth]{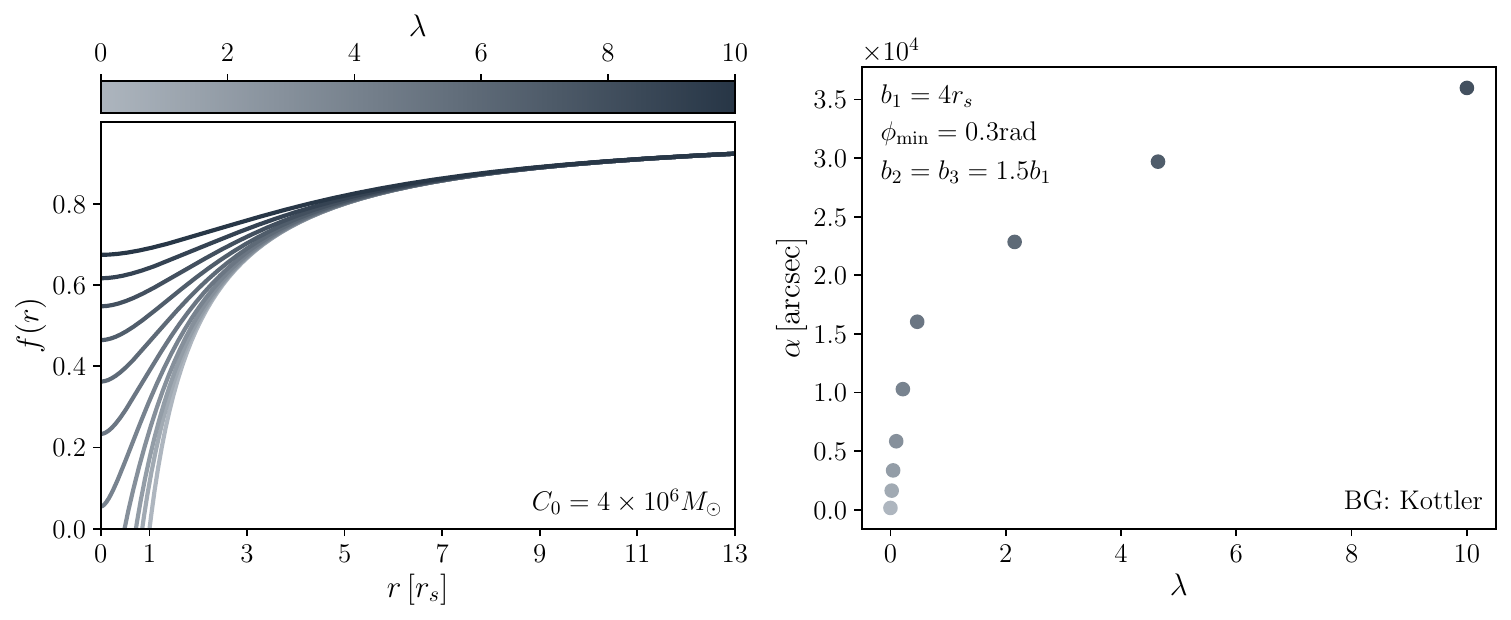}
    \caption{\textbf{Profiles and angular difference for a positive coupling constant.} The left panel illustrates the profiles of the metric function $f(r)$ for different values of $\lambda \in [0, 10]$, distinguished by their respective colors. The right panel shows the angular difference corresponding to these solutions when the Kottler solution is taken as the background (BG).}\label{fig:profiles_and_angular_diff}
\end{figure*}

\begin{figure*}
    \centering
    \includegraphics[width=0.9\linewidth]{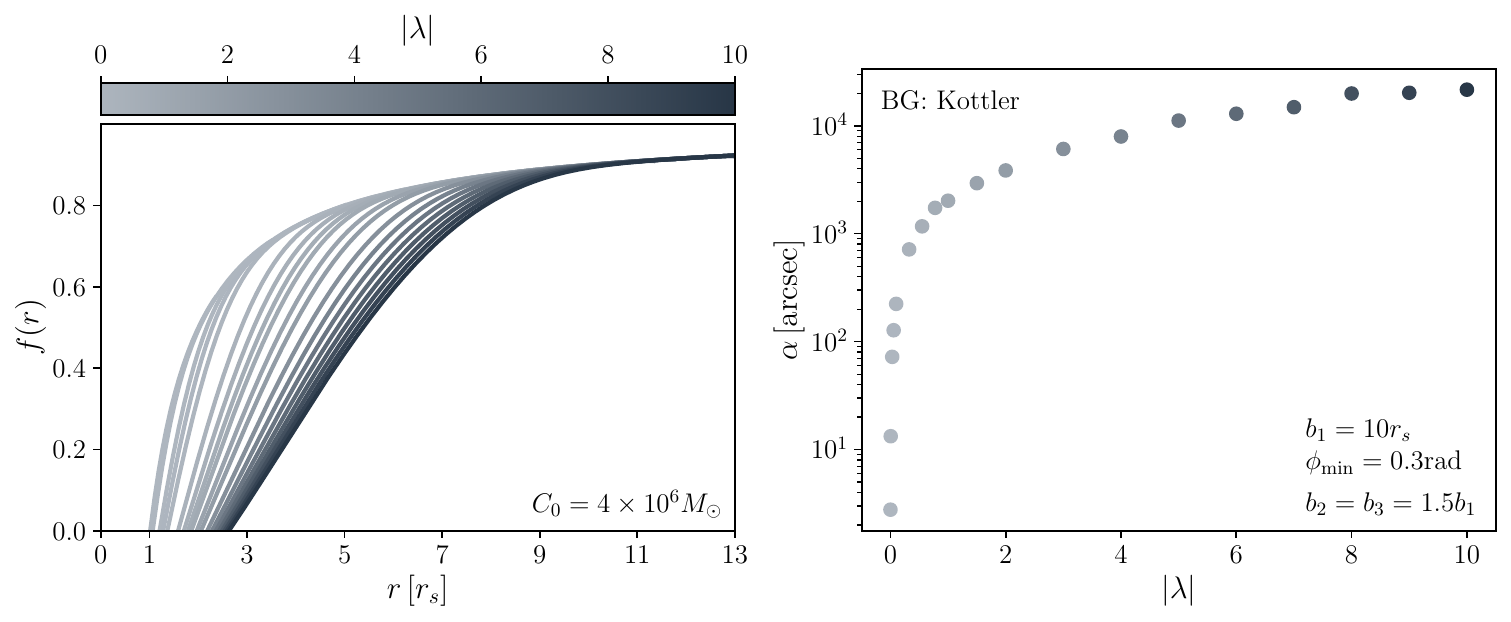}
    \caption{\textbf{Profiles and angular difference for a negative coupling constant.} Similar to Fig.~\ref{fig:profiles_and_angular_diff}, but in this case, $\lambda \leq 0$ and $b_1 = 10 r_s$.
 }\label{fig:profiles_and_angular_diff_N}
\end{figure*}

As pointed out at the beginning of this section, appropriately choosing the background spacetime makes it possible to disentangle the distinct contributions to the angular difference. Since our interest lies in identifying possible fingerprints of the cubic terms, we first consider the Kottler solution (with $M=C_0, \Lambda=\Lambda_{\text{CDM}}$) as the background in order to isolate their gravitational contribution arising from the black hole mass. Our second choice for the background is the Schwarzschild spacetime (with $M=C_0$). In this case, the contributions arise from both the mass and $\Lambda_{\text{Eff}}$. Fixing $\Lambda_{\text{Eff}} = \Lambda_{\text{CDM}}$ implies that the latter becomes relevant only when the triangular array is sufficiently far from the gravitational source, specifically when $b_1 > 10^{10} r_s$. Since we focused on the region where $b_1 < 10^{4} r_s$, the angular difference for both Kottler and Schwarzschild spacetimes remains the same (under double-precision floating-point numerical resolution). In other words, the contribution of $\Lambda_{\text{Eff}}$ to the cubic terms within this region is irrelevant compared to the mass contribution.

As initial examples, we considered as the spacetime of interest the (numerical and approximated) solutions of Einsteinian cubic gravity with the integration constant $C_0 = 4.3 \times 10^6 M_\odot$ (corresponding to the observed mass of SgrA${}^{*}$) and coupling constant values: $\lambda =-5.0, -1.5, 0.05, 0.1$. The impact parameters of the triangular array were chosen such that the relation $b_2 = b_3 = 1.5b_1$ holds, while $\phi_{\text{min}}$ was fixed at $0.3$ rad. Table~\ref{tabla_1} presents the angular difference $\alpha$ computed using the Kottler (and Schwarzschild) solution as the background. As noted above, in this scenario, the values of $\alpha$ encode the (gravitational) effect of the cubic terms due to the mass contribution. Our results indicate that as $|\lambda|\to0$ (for a fixed $b_1$), the value of $\alpha$ either diminishes or remains constant. This behavior highlights that the gravitational contribution of the cubic terms due to a massive source becomes negligible as $|\lambda|$ decreases. In the limit $\lambda = 0$, it is expected that $\alpha=0$. Consequently, the Kottler/Schwarzschild solution is recovered, corresponding to the regime where general relativity is restored. The cases where $\alpha$ remains constant despite the value of $\lambda$ correspond to triangular arrays that are sufficiently far from the gravitational source (e.g. $b_1 = 10^4 \, r_s$), such that the correction introduced by the coupling constant term in the asymptotic solution, Eq.~(\ref{Eq.SolAS}), becomes negligible. In contrast, the value of the angular difference increases considerably when the coupling constant $\lambda$ is fixed and the impact parameter $b_1$ decreases. This is a direct consequence of the triangular configurations being closer to the gravitational source, which makes the contribution of the cubic terms more relevant.

As a second example, we fixed the impact parameters $b_1=4r_s$, $b_2=b_3=1.5b_1$, and varied the coupling constant $\lambda$ within the interval $[0, 10]$ taking $C_0 = 4.3 \times 10^6 M_\odot$. The results are shown in Fig.~\ref{fig:profiles_and_angular_diff}. As observed from the left panel, when $\lambda \gtrapprox 0.1$, the solution does not present a black hole horizon (although the cosmological horizon remains present). This indicates that the value of $\lambda$ is constrained to $\lambda \lessapprox 0.1$ in order to ensure the presence of an exterior event horizon in black holes like SgrA${}^{*}$ (keeping in mind that the mass that we are using is very close to that of SgrA${}^{*}$). The right panel shows the angular difference for these solutions. As observed, this increases as $\lambda$ grows. However, the results seem to indicate that it asymptotically approaches a certain value. We do not delve deeper into identifying it, as it corresponds to non-physically viable values of $\lambda$ due to the large modifications to the spacetime metric near the massive source, leading even to the disappearance of the event horizon.

A similar analysis to the previous one is presented in Fig.~\ref{fig:profiles_and_angular_diff_N}, but in this case for $\lambda \in [-10, 0]$ and $b_1 = 10 r_s$. The last change is due to the exterior horizon moving to the right for this branch, and we need to ensure that the triangular array is not inside it. In this case, we always have an exterior event horizon (greater than $2C_0/M_{\textrm{Pl}}^2$), which could, in principle, reproduce black hole types like SgrA${}^{*}$. As observed from the left panel, when $\lambda$ decreases (increases in magnitude), the horizon's radius shifts to the right. However, this shift becomes progressively smaller; for example, the horizon displacement from a solution with $\lambda = -1$ to a solution with $\lambda = -2$ is approximately $0.2r_s$, while for $\lambda=-9$ to $\lambda=-10$ it is approximately $0.04r_s$. The right panel shows the angular difference. In this case, our results seem to indicate a linear increase (the $y$-axis is on a logarithmic scale). This is a direct consequence of the horizon approaching to the position of the triangular array. As discussed in the first example, this implies that the effect of the cubic terms becomes more relevant, and the differences relative to the triangular background array become more significant.

As a final example, we explored a solar scenario, considering a gravitational source with the configuration $C_0 = M_\odot$, $b_1 = R_\odot$, $\phi_{\text{min}} = 0.2$, and $b_2 = b_3 = 2R_\odot$. The angular difference obtained is approximately $\alpha_{\text{Sun}}\approx9.08 \times 10^{-9}$ arcsec. Although current technology does not yet allow us to measure such a small angular difference, this result opens up the possibility of experimentally detecting the contribution of the cubic terms in the future. This could be achieved using laser-beam baselines aboard missions such as LATOR~\cite{Turyshev:2003wt}, ASTROD-GW~\cite{Ni:2012eh}, and LISA~\cite{Bayle:2022hvs, Amaro_Seoane_2023} (see Table 1 in Ref.~\cite{Ni:2016wcv} for additional mission proposals). Such measurements could provide constraints on possible cubic modifications of general relativity.

From all the previous examples, we can conclude that the strongest effects of the cubic terms occur in the strong gravity regime (near the source).

\subsection{ Photon sphere}\label{Sec.III.C}
\begin{figure}
    \centering
    \includegraphics[width=0.9\linewidth]{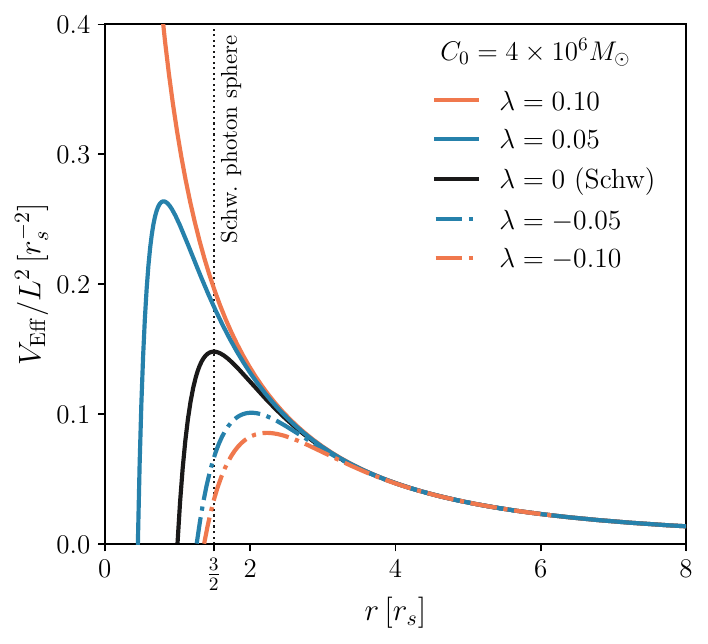}
    \caption{\textbf{Relationship between the effective potential $V_{\text{Eff}}$ and the coupling constant $\lambda$.} An increase in the value of $\lambda$ leads to a modification of $V_{\text{Eff}}$, causing its maximum to grow and shift to the left. For $\lambda\gtrapprox0.1$, the maximum disappears, resulting in the formation of a naked singularity. Note that for $\lambda<0$ ($\lambda>0$), the photon sphere is larger (smaller) than the Schwarzschild photon sphere ($\lambda=0$). The integration constant $C_0$ is fixed to the mass of SgrA$^{*}$.}\label{fig:EffPotentials}
\end{figure}

In Section~\ref{Sec.EvhPhS}, we identified the dependence of the horizon position on the sign of the coupling constant $\lambda$. (This result is validated numerically in Section~\ref{Sec.RegValit}.) The observational evidence suggests the presence of an event horizon in SgrA$^{*}$ which constrains the value of the coupling constant to $\lambda\lessapprox 0.1$. Moreover, since modifications to the black hole metric generically result in changes to the photon sphere and the black hole shadow, we foresee that another potential signature of cubic gravity can be found in the location of the photon sphere and the implications of its modification.

Using the fact that the weak equivalence principle holds in Einsteinian cubic gravity~(\ref{Eq.AECG}), and therefore test particles move along geodesics, we can derive the effective potential for the metric~(\ref{Eq.Metric}) following standard textbook calculations (see, e.g. Chapter $25$ of Ref.~\cite{Misner:1973prb} or Section $4.1.3$ of Ref.~\cite{Barranco:2021auj}), resulting in
\begin{equation}
V_{\text{Eff}}(r;L^2) = \frac{f(r) L^2}{r^2}.
\end{equation}
 Here, $L$ represents the angular momentum of the massless test particle. 

Figure~\ref{fig:EffPotentials} illustrates the relationship between the effective potentials (per unit squared angular momentum) and the coupling constant $\lambda$, with $C_0$ fixed to the mass of SgrA${}^{*}$. As expected, when $\lambda\neq0$, the potential profile undergoes changes compared to the Schwarzschild case ($\lambda=0$), with the potential barrier increasing and shifting to the left as the value of $\lambda$ grows. For $\lambda \gtrsim 0.1$, in the regime where naked singularities are present, the effective potential cannot reach zero, instead it grows indefinitely forming an infinite potential barrier. From the potential, we can infer the existence of a stable (unstable) circular orbit at a radius $r_0$ where $V_{\text{Eff}}(r_0)=E$ (total energy), $V_{\text{Eff}}'(r_0)=0$ and $V_{\text{Eff}}''(r_0)>0$ ($V_{\text{Eff}}''(r_0)<0$). Directly from Fig.~\ref{fig:EffPotentials}, we can conclude that for $\lambda > 0.1$, a photon sphere does not exist, which contradicts observational evidence. In addition, even for $|\lambda|<0.1$ we observe significant changes (around $50\%$) in the location of the photon sphere with respect to that of a Schwarzschild black hole. This may lead to sizable changes in the shadow radius of these black holes. For instance, for a Schwarzschild black hole, the shadow radius as seen by an observer in the asymptotically flat region is simply proportional to the radius of the photon sphere. For more general asymptotically flat black holes the relation is different, but a similar relative change is expected (see, e.g. Ref.~\cite{Vertogradov:2024dpa} for a general approximate treatment and Red.~\cite{Gutierrez-Cano:2024oon} for a particular case in Einsteinian cubic gravity coupled to non-linear electrodynamics). A detailed analysis of these changes in our non-asymptotically flat spacetimes is left for future work.  

\section{Conclusions}\label{Sec.Concl}

In this paper, we have analyzed Einsteinian cubic gravity in the strong gravity regime. This theory is described by the Einstein-Hilbert action with a constant $\Lambda$ plus a specific non-topological (in four dimensions) cubic term arising from contractions of the Riemann tensor, which modifies the spacetime dynamics while preserving the same spectrum as general relativity. Using a static, spherically symmetric Ansatz (characterized by a single $f(r)$ function) for the spacetime, we derived the dynamical field equations in terms of an integration constant $C_0$.

In the first part of the paper, we studied exact and approximate solutions, establishing the existence of maximally symmetric de Sitter solutions with an effective cosmological constant determined both by $\Lambda$ and by the coupling constant $\lambda$. When maximal symmetry is abandoned, the structure of the solutions in cubic gravity is no longer the same as in general relativity, e.g. instead of having Schwarzschild or Kottler as solutions, we have more general functional forms of the metric that are already identified in the weak coupling, asymptotic and near horizon approximations.
These new solutions generalize previous findings reported in the literature by incorporating the constant $\Lambda$ and expressing them in terms of a single integration constant, $C_0$.

In the weak coupling regime, we identified an iterative solution whose deviations from Kottler and Schwarzschild are weighted by the dimensionless parameter $\lambda\Lambda^2/M_{\text{Pl}}^5$. Furthermore, the Kottler mass and cosmological constant get redefined by the contribution coming from the cubic terms. Similar findings are unveiled by using an asymptotic (instead of weak coupling) approximation. After verifying that we can reproduce results reported in the literature, where the constant $\Lambda$ is treated as the observational cosmological constant, we took a different approach by considering $\Lambda_{\text{Eff}}$ as the observational cosmological constant. We justify this choice by noting that the scalar curvature due to the black hole is dictated by $\Lambda_{\rm{Eff}}$ and, if the black hole is embedded in a large-scale de Sitter cosmological metric, then this curvature must agree with the curvature of the cosmological model, which is determined by the observational value of the cosmological constant.

In the strong gravity regime, we derived a near-horizon iterative solution expressed in terms of the integration constant $C_0$ and the second derivative of the metric function at the horizon, $f''(r_h)$. From our solution, it is possible to recover the horizon properties reported in the literature for the specific cases $\Lambda=\lambda=0$ and $\Lambda=0, \lambda<0$. We also discuss the case $\Lambda=0, \lambda>0$, giving an original explanation for the properties of the horizon. Finally, we discuss the more general situation $\Lambda\neq0, \lambda\neq0$, where using the full numerical solution we identify that a negative (positive) coupling constant leads to exterior horizons that are larger (smaller) than $2C_0/M_{\textrm{Pl}}^2$. Furthermore, we found that the near-horizon approximated solution is ruled out when $\lambda>0$, but remains valid for $\lambda<0$ within a small region near the horizon. The asymptotic and weak coupling solutions satisfactorily reproduce the full numerical solution for regions far from the horizon.

In the second part of the paper, we have discussed potential signatures of Einsteinian cubic gravity in the strong gravitational regime, focusing on the modifications introduced by the cubic terms in light deflection compared to the predictions of general relativity. In particular, we analyzed the angular difference (previously presented in Ref.~\cite{S_nchez_2023}) and discussed potential changes of the photon sphere associated with these solutions. Using the angular difference with either the Kottler or Schwarzschild solution as the background, we isolated the gravitational contribution of the cubic terms and identified that their signatures are important in regions close to the massive source. The idea of using the angular difference measurement in the Solar System was explored as a final example. In this case, the cubic contribution was on the order of nano-arcseconds. Although this resolution is currently beyond experimental capabilities, the use of the Solar System as a laboratory for testing modified gravity theories cannot be entirely ruled out. For instance, future experiments such as LISA are expected to achieve resolutions around a micro-arcsecond. While this does not yet reach the precision needed to detect such small angular differences, it opens the door to the possibility of observing signals of modified gravity in upcoming high-precision measurements.

From the study of the effective potential and its implications for the associated photon sphere in solutions with $C_0$ fixed to the SgrA$^{*}$ mass, we identified that the coupling constant is constrained to $\lambda \lessapprox 0.1
$. Additionally, we observed that as the value of $\lambda$ increases, the potential barrier grows and shifts to the left. Significant changes were also noted in the location of the potential maximum (photon sphere position) compared to the Schwarzschild case, potentially leading to substantial deviations in the shadow radius of these black holes relative to those predicted by Schwarzschild black holes. These findings underscore the potential of the cubic terms to influence observable phenomena, such as black hole shadows. This idea will be explored in future works, as well as the influence of the structural shape of the triangular configuration on the angular difference in a Solar System scenario.

\begin{acknowledgments}
A.A.R. acknowledges funding from a postdoctoral fellowship from ``Estancias Posdoctorales por México para la Formación y Consolidación de las y los Investigadores por México''.
F.S. is supported by ``Becas Nacionales para Estudios de Posgrados" and J.C. by CONACyT/DCF/320821.

\end{acknowledgments}

\appendix

\section{Complementary material}\label{Ap.CompMat}

In this appendix, we present the complete equations that were shown in abbreviated or partial form in the main text, as well as a complementary discussion on the maximally symmetric solutions.

\subsection{Vacuum covariant field equations}\label{Ap.VCMEq}

In Section~\ref{Sec.TS}, the vacuum covariant field equations~(\ref{Eq.MotCov}), were presented, where the modifications introduced by the cubic terms are represented by the tensor  $\mathcal{P}_{\mu\nu}$, defined as follows:
\begin{widetext}
\begin{align}\label{A.Eq.MotCov}
\mathcal{P}_{\mu\nu}=&g_{\mu\nu}A-24 R_{\mu\nu} \Box R +48 R_{\mu}{}^{\rho}A_{\nu\rho}+48\nabla_{\nu}R_{\sigma\rho}(\nabla^{\rho}R_{\mu}{}^{\sigma}-\nabla_{\mu}R^{\sigma\rho})+R_{\mu\sigma\nu\rho}(96\Box R^{\sigma\rho}-72\nabla^{\rho}\nabla^{\sigma}R)-24 R^{\sigma\rho}A_{\sigma\mu\rho\nu}\nonumber\\
&-144\nabla_{\rho}R_{\mu\gamma\nu\sigma}\nabla^{\gamma}R^{\sigma\rho}+48(R_{\nu\rho\sigma\gamma}-3R_{\nu\sigma\rho\gamma})\nabla^{\gamma}\nabla^{\rho}R_{\mu}{}^{\sigma}+48(2\nabla^{\gamma}\nabla_{\nu}R^{\sigma\rho}-3\nabla^{\gamma}\nabla^{\rho}R_{\nu}{}^{\sigma})R_{\mu\sigma\rho\gamma}+48(R_{\nu}{}^{\sigma}R^{\rho\gamma}\nonumber\\
&+\nabla^{\gamma}\nabla^{\rho}R_{\nu}{}^{\sigma})R_{\mu\rho\sigma\gamma}-12 R_{\mu}{}^{\sigma\rho\gamma}B_{\nu\rho\sigma\gamma}-24(\nabla_{\mu}R_{\nu\sigma}+\nabla_{\nu}R_{\mu\sigma}+2\nabla_{\sigma}R_{\mu\nu})\nabla^{\sigma}R-144(\nabla_{\sigma}R_{\nu\rho}-\nabla_{\rho}R_{\nu\sigma})\nabla^{\rho}R_{\mu}{}^{\sigma}\nonumber\\
&+48R_{\nu}{}^{\sigma}(\Box R_{\mu\sigma}-R_{\mu}{}^{\rho\gamma\eta}R_{\sigma\gamma\rho\eta})+24(\nabla_{\sigma}R_{\nu\eta\rho\gamma}-6\nabla_{\gamma}R_{\nu\rho\sigma\eta})\nabla^{\eta}R_{\mu}{}^{\sigma\rho\gamma}-48(\nabla_{\mu}R_{\nu\sigma\rho\gamma}+\nabla_{\nu}R_{\mu\sigma\rho\gamma}+3\nabla_{\rho}R_{\mu\sigma\nu\gamma}\nonumber\\
&-4\nabla_{\gamma}R_{\mu\sigma\nu\rho})\nabla^{\gamma}R^{\sigma\rho}+12 B_{\mu\nu}.
\end{align}
with
\begin{subequations}
\begin{align}\label{A.Eq.A}
A=&15R^3+R\left(9R_{\mu\nu\rho\sigma}^2-108 R_{\mu\nu}^2\right)+8R^{\mu\nu}\left(22R_{\mu}{}^{\rho}R_{\nu\rho}+9\nabla_\nu\nabla_{\mu}R-6\Box R_{\mu\nu}\right)-2(7R_{\mu\nu}{}^{\rho\sigma}R^{\mu\nu\gamma\beta}R_{\gamma\beta\rho\sigma}-6(\nabla_{\mu}R)^2\nonumber\\
&-72\nabla_{\mu}R_{\nu\rho}\nabla^{\rho}R^{\nu\mu}+48(\nabla_{\mu}R_{\nu\rho})^2+24R_{\mu\nu\rho\gamma}\nabla^{\gamma}\nabla^{\nu}R^{\mu\rho}),\\
A_{\nu\rho}=&\Box R_{\nu\rho} + R_{\nu}{}^{\sigma}R_{\rho\sigma}+R^{\sigma\gamma}R_{\nu\sigma\rho\gamma}-R_{\nu}{}^{\sigma\gamma\beta}R_{\rho\gamma\sigma\beta},\\
A_{\sigma\mu\rho\nu}=& 10R_{\sigma}{}^{\gamma}R_{\mu\rho\nu\gamma}-R_{\mu\sigma}{}^{\gamma\beta}R_{\nu\rho\gamma\beta}+2(R_{\mu}{}^{\gamma}{}_{\sigma}{}^{\beta}R_{\nu\beta \rho \gamma} - R_{\mu}{}^{\gamma}{}_{\nu}{}^{\beta} R_{\sigma\gamma \rho\beta}-2\nabla_{\nu}\nabla_{\mu}R_{\sigma\rho}+2\nabla_{\rho}\nabla_{\mu}R_{\nu\sigma}+2\nabla_{\rho}\nabla_{\nu}R_{\mu\sigma}\nonumber\\
&+\nabla_{\rho}\nabla_{\sigma}R_{\mu\nu}+\Box R_{\mu\sigma\nu\rho}),\\
B_{\mu\nu}=&\nabla_{\mu}R\nabla_{\nu}R+4\nabla_{\mu}R_{\sigma\rho}\nabla^{\rho}R_{\nu}{}^{\sigma}+8R_{\nu\sigma\rho\gamma}\nabla^{\gamma}\nabla_{\mu}R^{\sigma\rho}+12R^{\sigma\rho\gamma\eta}\nabla_{\eta}\nabla_{\rho}R_{\mu\sigma\nu\gamma},\\
B_{\nu\rho\sigma\gamma}=&R_{\nu}{}^{\beta}{}_{\sigma}{}^{\eta}(R_{\rho\gamma\beta\eta}-6R_{\rho\beta\gamma\eta})-4R_{\nu}{}^{\beta}{}_{\rho}{}^{\eta}(2R_{\sigma\beta\gamma\eta}+3R_{\sigma\eta\gamma\beta}) + R_{\nu\rho}{}^{\beta\eta}(R_{\sigma\gamma\beta\eta}-18R_{\sigma\beta\gamma\eta})-R_{\nu\sigma}{}^{\beta\eta}R_{\rho\gamma\beta\eta},
\end{align}
\end{subequations}
\end{widetext}
where the square of a tensor denotes its self-contraction, e.g. $R_{\mu\nu}^2:=R_{\mu\nu}R^{\mu\nu}$, and $\Box:=\nabla_{\mu}\nabla^{\mu}$ represents the d’Alembert operator.

\subsection{Maximally symmetry solutions}\label{Ap.SecMSS}

In Section~\ref{Sec.ExSol}, we demonstrated the existence of maximally symmetric de Sitter-like solutions in Einsteinian cubic gravity. For real and positive (negative) $\Lambda_{\text{Eff}}$ roots of the equation:
\begin{align}\label{A.LambEffEq}
\Lambda_{\text{Eff}}^3 + \frac{M_{\textrm{Pl}}^5}{16\lambda}\Lambda_{\text{Eff}} - \frac{M_{\textrm{Pl}}^5\Lambda}{48\lambda} = 0,
\end{align}
the solutions exhibit behavior corresponding to de Sitter (anti-de Sitter) solutions. Next, we present the real parameter space corresponding to the roots of the equation above. 

Fixing the $\Lambda_{\text{Eff}}$ value to the observational cosmological constant, we arrive at the linear relation
\begin{align}\label{A.Eq.ConstRelat}
\Lambda=\left(3.32+\frac{6.48\times 10^{-103}\hbar^5}{M_{\textrm{Pl}}^5 c^5}\lambda\right)\times 10^{-52} m^{-2},
\end{align}
where the physical units are restored, and $M_{\textrm{Pl}}, \hbar$, and $c$ are expressed in SI units, and  $\lambda$ in inverse meters. Notice that for an Einstein cubic model without a cosmological constant (i.e. $\Lambda=0$ in Eq.~(\ref{Eq.AECG})), the observational cosmological constant is recovered only for a negative coupling constant, but not for a positive one. For models with $\Lambda\neq0$, it is possible to recover the observational value within the parameter space where Eq.~(\ref{A.Eq.ConstRelat}) holds.

\subsection{Weak coupling solution}\label{A.WCS}

In Section~\ref{Sec.WCS}, the weak coupling solution for Eq.~(\ref{Eq.System2}) is obtained iteratively. The functions $h_1$, $h_2$, and $h_3$ are presented in Eq.~(\ref{Eq.WcsCoeff}). The full expression for the $H_3$ term in $h_3$ is:
\begin{widetext}
\begin{align}
H_3=&-\frac{10935 C_0}{14\Lambda ^3 M_{\textrm{Pl}}^2 r^7} \bigg(1+\frac{3699 C_0}{35M_{\textrm{Pl}}^2r}-\frac{1411589 C_0^2}{5670 M_{\textrm{Pl}}^4 r^2}-\frac{10797 C_0}{28\Lambda M_{\textrm{Pl}}^2 r^3}+\frac{12158 C_0^2}{7\Lambda M_{\textrm{Pl}}^4 r^4}+\frac{3078 C_0}{7\Lambda ^2 M_{\textrm{Pl}}^2 r^5}-\frac{102204 C_0^2}{35\Lambda^2 M_{\textrm{Pl}}^4 r^6}-\frac{5466863 C_0^3}{2835\Lambda M_{\textrm{Pl}}^6 r^5}\nonumber\\
&+\frac{179061 C_0^3}{28 \Lambda^2 M_{\textrm{Pl}}^6 r^7}-\frac{4372576 C_0^4}{945\Lambda^2 M_{\textrm{Pl}}^8 r^8}\bigg).
\end{align}
\end{widetext}

\subsection{Asymptotic solution}\label{A.AS}

In Section~\ref{Sec:AsymSol}, we obtain the iterative asymptotic solution for Eq.~(\ref{Eq.System2}), showing the main terms in Eq.~(\ref{Eq.SolAS}). Next, we present the corresponding terms for the first ten terms:
\begin{widetext}
\begin{align}\label{A.Eq.SolAS}
    f_{\text{asy}}(r)=&1-\frac{\Lambda_{\text{Eff}}}{3} r^2-\frac{6M_{\textrm{Pl}}^3}{3 M_{\textrm{Pl}}^5+16\lambda \Lambda_{\text{Eff}}^2}\frac{C_0}{r}+\frac{432\lambda}{M_{\textrm{Pl}}^9\left(1+\frac{16\lambda \Lambda_{\text{Eff}}^2}{3M_{\textrm{Pl}}^5}\right)^3}\left(1-\frac{14 \Lambda_{\text{Eff}}}{27}r^2\right)\frac{C_0^2}{r^6}\nonumber\\
    &+\frac{1202688\lambda^2 \Lambda_{\text{Eff}}}{M_{\textrm{Pl}}^{16}\left(1+\frac{16\lambda \Lambda_{\text{Eff}}^2}{3M_{\textrm{Pl}}^5}\right)^5}\left(1-\frac{1262 \Lambda_{\text{Eff}}}{7047}r^2-\frac{23 M_{\textrm{Pl}}^5}{37584\lambda \Lambda_{\text{Eff}}}r^2\right)\frac{C_0^3}{r^9}+\mathcal{O}\left(\lambda^2, \frac{C_0^4}{r^{10}}\right),
\end{align}
\end{widetext}
where $\Lambda_{\text{Eff}}$ is the effective cosmological constant and $C_0$ is an integration constant.

\bibliography{ref}

\end{document}